\documentclass[aps,twocolumn,twoside,secnumarabic,balancelastpage,amsmath,amssymb,hyperref=pdftex,superscriptaddress 
]{revtex4}
\usepackage{graphics}      
\usepackage[pdftex]{graphicx}      
\usepackage{longtable}     
\usepackage{bm}            
\usepackage{verbatim}
\usepackage{mathtools}
\usepackage[colorlinks=true]{hyperref}
\usepackage{epstopdf}

\begin{document}

\title{Weak phonon-mediated pairing in BiS$_2$ superconductor from first principles}
\date{\today}

\author{Corentin Morice}
\affiliation{Cavendish Laboratory, University of Cambridge, Cambridge CB3 0HE, United Kingdom}
\author{Ryosuke Akashi}
\affiliation{University of Tokyo, 7-3-1 Hongo, Bunkyo-ku, Tokyo 113-
8656, Japan}
\author{Takashi Koretsune}
\affiliation{Center for Emergent Matter Science, RIKEN, Wako, Saitama 351-0198, Japan}
\author{Siddharth S. Saxena}
\affiliation{Cavendish Laboratory, University of Cambridge, Cambridge CB3 0HE, United Kingdom}
\author{Ryotaro Arita}
\affiliation{Center for Emergent Matter Science, RIKEN, Wako, Saitama 351-0198, Japan}

\begin{abstract}
Superconductivity in novel bismuth-sulphur superconductors has attracted large research efforts, both experimental and theoretical, but a consensus on the nature of superconductivity in these materials has yet to be reached. Using density functional theory for superconductors, we study the electron-phonon pairing mechanism in LaO$_{0.5}$F$_{0.5}$BiS$_2$. We first confirm the presence of a commensurate charge density wave instability, in accordance with previous studies. Using a recently developed integration scheme for the electron-phonon coupling, we found that its strength is much lower than previously calculated, due to improved density of state calculations. We finally conclude that conventional phonon-mediated pairing cannot explain the high superconducting transition temperatures observed in this material.
\end{abstract}

\maketitle

\section{Introduction}

Superconductivity was discovered in two layered materials sharing the same layer of bismuth and sulphur atoms \cite{Mizuguchi2012, Mizuguchi2012a, Awana2013}, which has been shown to be the superconducting layer \cite{Usui2012, Usui2015, Morice2015}. The material having reached the highest superconducting transition temperature ($T_c$) of 10.6 K so far is La(O,F)BiS$_2$ \cite{Mizuguchi2012a}. Its parent compound is insulating and becomes superconducting upon doping, which raises the Fermi level in a rigid band fashion \cite{Usui2012}. Various variations of this composition have been attempted, of which the most studied have been the replacement of lanthanum with a different lanthanide \cite{Xing2012, Jha2013c, Demura2013, Jha2013a, Zhai2014, Yazici2013b}.

Interest in this family of superconductors was sparked by the theoretical prediction of the presence of a charge density wave instability in La(O,F)BiS$_2$ \cite{Wan2013, Yildirim2013}, and its observation in EuFBiS$_2$ \cite{Zhai2014}. Moreover, ferromagnetism was measured in the superconducting phase of Ce(O,F)BiS$_2$, in agreement with ab-initio calculations \cite{Xing2012, Morice2016}. Finally, quantum critical fluctuations of the magnetic moments were measured in CeOBiS$_2$ \cite{Higashinaka2015}.

La(O,F)BiS$_2$ diplays a large range of superconducting transition temperatures between 2.5 and 11.5 K \cite{Mizuguchi2014a, Zhang2016}. Its change with doping was measured, and gives a dome structure centred on $x=0.5$ \cite{Deguchi2013}. The synthesis method has been shown to be crucial in changing $T_c$: the crystals synthesised using a simple solid state reaction at ambient pressure (AP) have $T_c \sim 3$ K, while the ones having gone through an extra high pressure annealing stage (HP) have $T_c \sim 10$ K \cite{Mizuguchi2012a}. Interestingly, this change is revertible: HP samples annealed again at ambient pressure recover the structure and properties of AP samples \cite{Kajitani2014}. The difference in crystal structure between AP and HP samples is hard to characterise, as X-ray peaks of HP samples are much broader than the ones of AP samples \cite{Mizuguchi2012a, Deguchi2013}. This difference has however been linked to strain along the c-axis \cite{Kajitani2014}, and to a change from a larger positive Hall resistance in AP samples to a smaller negative one in HP samples \cite{Pallecchi2014}.

The evolution of La(O,F)BiS$_2$ under pressure has been measured for both AP and HP samples. $T_c$ in AP samples exhibit a sharp increase under pressure from around 3 K to around 9.5 K at 0.7 GPa \cite{Wolowiec2013, Tomita2014}. In comparison, in HP samples, $T_c$ rises very slightly with pressure \cite{Kotegawa2012, Wolowiec2013}. This change in $T_c$ has been shown to coincide with a structural phase transition from tetragonal to monoclinic \cite{Tomita2014}. However relating this to the crystal structure of the HP samples is difficult since X-ray and neutron measurements both gave a tetragonal structure for HP samples \cite{Mizuguchi2012a,Lee2013}

The prediction of a charge density wave instability in La(O,F)BiS$_2$ \cite{Wan2013,Yildirim2013} led to in-depth studies of its vibrational modes using neutron diffraction \cite{Lee2013, Athauda2015}. No phonon anomaly in either doping or temperature change was detected, which suggested that the electron-phonon coupling could be much weaker than the calculated values \cite{Lee2013}. Furthermore, neutron diffraction data yielded the presence of local charge fluctuations \cite{Athauda2015}.

Angle-resolved photoemission spectroscopy (ARPES) was performed on La(O,F)BiS$_2$, Ce(O,F)BiS$_2$ and Nd(O,F)BiS$_2$ \cite{Zeng2014, Ye2014, Terashima2014, Sugimoto2015}. Large discrepancies with electronic structure calculations were obtained, particularly the absence of the central nested Fermi pocket \cite{Zeng2014, Ye2014, Sugimoto2015}. This was attributed to a level of doping lower than expected \cite{Ye2014, Sugimoto2015}. This was confirmed by measurements on precisely calibrated samples \cite{Terashima2014}. Finally, electronic correlations were found to be weak in these systems \cite{Ye2014, Terashima2014}

The diversity in the phenomenology of these materials drove theoretical studies to probe the possibility of unconventional pairing. Spin-fluctuation mediated pairing has been calculated using the random-phase approximation, \cite{Martins2013, Zhou2013, Wu2014}, numerics \cite{Liang2014, Liu2014} and functional renormalisation group \cite{Yang2013}.

Ab-initio calculations outputted large electron-phonon couplings, between 0.8 and 0.85 \cite{Wan2013, Yildirim2013, Li2013}. Two calculations found unstable phonon modes around ($\pi$,$\pi$,0) \cite{Wan2013, Yildirim2013}, leading to a commensurate charge density wave \cite{Wan2013, Yildirim2013}. The phonon dispersion along the $c$ axis was found to be small, highlighting the small magnitude of the interlayer coupling across the buffer layers \cite{Wan2013}. All these studies concluded that superconductivity in La(O,F)BiS$_2$ is strongly coupled and conventional \cite{Wan2013, Yildirim2013, Li2013}. The same conclusion has been reached in La(O,F)BiSe$_2$ \cite{Feng2014a}

In this paper, we study conventional superconductivity in LaO$_{0.5}$F$_{0.5}$BiS$_2$, using density functional theory for superconductors \cite{Luders2005,Marques2005a}. We first confirm the arising of a commensurate charge density wave instability, in accordance with previous studies \cite{Wan2013, Yildirim2013}. Using a novel integration scheme for the electron-phonon coupling, we find that its strength is much lower than previously calculated, due to our improved convergence of density of state calculations. We finally conclude that conventional phonon-mediated pairing cannot explain the high $T_c$s observed in this material.

\section{Methods}

Electronic structures, dynamical matrices and electron-phonon coupling coefficients were calculated using norm-conserving pseudopotentials and a plane wave basis set as implemented in {\sc quantum espresso} \cite{Giannozzi2009}. The plane-wave energy cutoff for the wavefunctions was set to 90 Ry. We also used the all-electron full-potential linearised augmented plane-wave code Elk \cite{Elk}. Electron dielectric functions were calculated within the frequency-dependent random-phase approximation.

We employed the exchange-correlation potential implemented by Perdew and Zunger based on the local density approximation \cite{Perdew1981}. The relativistic effects are mostly treated at the scalar-relativistic level \cite{Koelling1977a}, whereas the spin-orbit coupling was explicitly included for re-examining the charge instability (see below).

We used density functional theory for superconductors to calculate superconducting transition temperatures ($T_c$) from first principles \cite{Luders2005,Marques2005a}. It is an ab-initio theory which allows one to calculate the conventional strong-coupling $T_c$ within the level of the Migdal-Eliashberg theory \cite{Migdal1958, Eliashberg1960, Eliashberg1961} and without ajustable parameters.

It consists in solving self-consistently the gap equation:
\begin{equation*}
\Delta_{n\textbf{k}} = -\mathcal{Z}_{n\textbf{k}} \Delta_{n\textbf{k}} - \frac{1}{2} \sum_{n'\textbf{k}'} \mathcal{K}_{n\textbf{k}n'\textbf{k}'} \frac{\tanh[(\beta/2) E_{n'\textbf{k}'}]}{E_{n'\textbf{k}'}} \Delta_{n'\textbf{k}'}
\end{equation*}
where $n$ and $n'$ run over bands, $\textbf{k}$ and $\textbf{k}'$ over crystal momentum, $\beta$ is the inverse temperature, $\Delta_{n\textbf{k}}$ is the superconducting gap function, and $E_{n\textbf{k}} = \sqrt{(\epsilon_{n\textbf{k}} - \mu )^2 + \Delta_{n\textbf{k}}^2}$ where $\epsilon_{n\textbf{k}}$ is the Kohn-Sham one-particle energy and $\mu$ is the chemical potential for the normal state. In subsequent calculations, $\mu$ is approximated to the Fermi level for the normal state $E_F$, where the temperature dependence is ignored. $\mathcal{Z}$ and $\mathcal{K}$ are exchange-correlation kernels. $\mathcal{Z}$ represents the mass renormalisation of the normal-state band structure due to phonon exchange, while $\mathcal{K} = \mathcal{K}^{\text{ph}} + \mathcal{K}^{\text{el}}$ represents electron-phonon and electron-electron interactions.

This gap equation is derived by using an extended version of the standard finite-temperature density-functional procedure \cite{Mermin1965} consisting in minimising a grand canonical potential function of three densities: the density of electrons, of nuclei, and the superconducting order parameter \cite{Luders2005}. Choosing a system of non-interacting electrons but interacting nuclei as the Kohn-Sham system, one obtains the Kohn-Sham equations in the Bogoliubov-de Gennes form. Neglecting the feedback effect of the superconducing gap on the normal electronic states finally gives the gap equation \cite{Luders2005}.

\begin{figure}
\centering
\includegraphics[width=9.5cm]{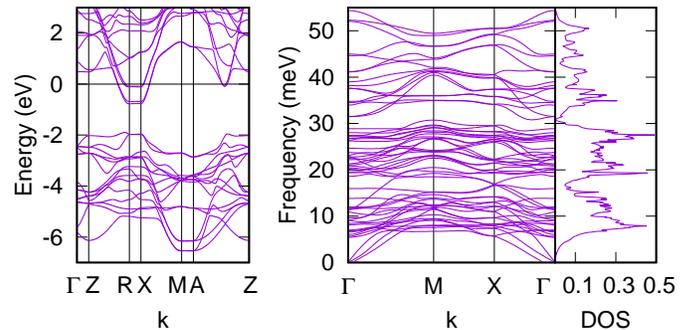}
\caption{Electronic band structure in the original unit cell (left), and phonon dispersion and density of states (DOS) in the $\sqrt{2}\times\sqrt{2}$ structure (right).}
\label{dispersion}
\end{figure}

The kernels of the gap equation, defined by the previous procedure as functional derivatives of terms of the grand canonical potential, are derived perturbatively using many-body theory \cite{Gorling1994, Luders2005}. We employed the phononic kernels $\mathcal{K}^{\text{ph}}$ and $\mathcal{Z}^{\text{ph}}$ averaged over $k$-point defined in equation (23) in \cite{Marques2005a} and equation (40) in \cite{Akashi2013a}, respectively. The phonon and electron-phonon coupling properties were calculated using density functional perturbation theory \cite{Baroni2001, Giannozzi2009}. The pairing and mass-renormalisation effects are thus treated at the level of the Migdal-Eliashberg theory with varying density of states \cite{Pickett1982}.

The electronic kernel $\mathcal{K}^{\text{el}}$ is calculated using the frequency-dependent dielectric function \cite{Akashi2013, Akashi2013c, Akashi2015}, thus including dynamical Coulomb effects \cite{Takada1978}.

\section{Results}

\subsection{Structural instability}

We calculated the phonon dispersion in the tetragonal unit cell using density functional perturbation theory on a sampling of the Brillouin zone of $4 \times 4 \times 2$ $\textbf{q}$-points with {\sc quantum espresso}. Hereafter we refer to the phonon wave number by $\textbf{q}$. The obtained dispersion features a structural instability which extends all along the $\Gamma$-M line, matching previous studies.

In order to test the robustness of this instability, both to other approximations and to the inclusion spin-orbit coupling, we recalculated this dispersion using an all-electron calculation with and without spin-orbit coupling, and a $\textbf{q}$-grid of $2 \times 2 \times 1$ points. This yielded two dispersions, which are similar, with a slightly reduced depth of the imaginary frequencies from $-20$ meV to $-14$ meV. This confirms that this instability is robust.

\begin{figure}
\centering
\includegraphics[width=8cm]{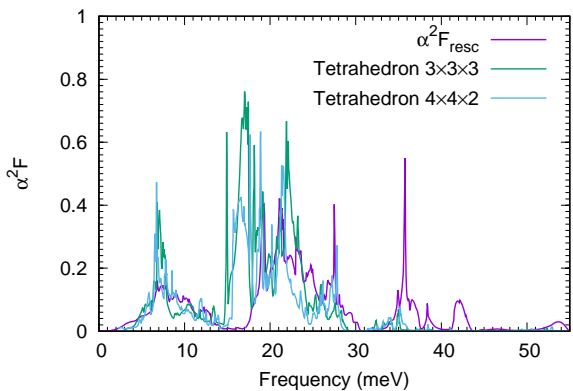}
\caption{Eliashberg functions calculated using the rescaling scheme ($\alpha^2 F_{\text{resc}}$) for a smearing $\approx 0.068$ eV, and the tetrahedron scheme with two different $\textbf{q}$-point grids, both shifted along the all the axes. The latter is $\alpha^2 F_{\text{tetra}}$. The corresponding values of $\lambda$ are 0.44, 0.58 and 0.47.}
\label{a2F-figure}
\end{figure}

We obtained the fully relaxed structure by using a unit cell of dimensions $\sqrt{2} a \times \sqrt{2} a$ where $a$ is the unit cell parameter of the tetragonal structure. For convenience we call this new crystal structure the $\sqrt{2} \times \sqrt{2}$ structure.

We calculated the phonon dispersion in this $\sqrt{2} \times \sqrt{2}$ structure  (Figure \ref{dispersion}). The obtained dispersion does not feature the instability \footnote{We find imaginary modes in the small region around the Gamma point, which could be due to insufficient energy cutoff for charge density.}. We therefore use this structure to calculate electron-phonon coupling coefficients.

\subsection{Electron-phonon coupling}

The effect of electron-phonon coupling on conventional superconductivity is well represented by the Eliashberg function \cite{SuperconductivityVol1}:
\begin{align}
\alpha^2 F (\omega) = &\frac{1}{N(E_F)} \sum_{\nu \textbf{q}} \sum_{n n' \textbf{k}} |g^{\nu \textbf{q}}_{n \textbf{k}+\textbf{q}, n' \textbf{k}}|^2 \delta (\epsilon_{n\textbf{k}+\textbf{q}} - E_F) \nonumber \\ &\times \delta (\epsilon_{n'\textbf{k}} - E_F) \delta (\omega - \omega_{\nu \textbf{q}})
\label{a2F}
\end{align}
where $\omega$ is the frequency, $n$ and $n'$ run over electronic bands, $\nu$ over phononic bands, $\textbf{k}$ and $\textbf{q}$ over the Brillouin zone, and $g$ are the electron-phonon coupling coefficients. Integrating the Eliashberg function over frequency gives the electron-phonon coupling factor $\lambda$:
\begin{equation}
\lambda = 2 \int d\omega \frac{\alpha^2 F}{\omega}
\label{lambda}
\end{equation}
The characteristic phonon frequency $\omega_{ln}$ is defined by:
\begin{equation}
\omega_{ln} = \exp \left( \frac{\int d\omega \frac{\alpha^2 F (\omega)}{\omega} \ln (\omega)}{\int d\omega \frac{\alpha^2 F (\omega)}{\omega}} \right)
\label{McMillan-Allen-Dynes}
\end{equation}
The effects of the electron-phonon coupling on $T_c$ are well quantified by the McMillan-Allen-Dynes formula \cite{McMillan1968, Allen1975}:
\begin{equation}
T_c = \frac{\omega_{ln}}{1.2} \exp \left( -\frac{1.04 (1+\lambda)}{\lambda - \mu^*(1+0.62\lambda)} \right)
\end{equation}
where $\mu^*$ is an adjustable pseudopotential parameter representing the effective electron-electron Coulomb repulsion suppressing the pairing instability \cite{Morel1962}.

In practice, the integration of equation \ref{a2F} is carried out by approximating the singular delta function by a smeared function $\tilde{\delta}$ such as the Hermite-Gaussian \cite{Methfessel1989}. However, this treatment can introduce two types of quantitative errors. First, the integrated value converged with respect to the $k$-point density suffers from a systematic error due to the width of the smearing function. Second, the convergence with respect to the $k$-point density tends to be very difficult to achieve, especially when the smearing width is small.

\begin{figure}[t]
\centering
\includegraphics[width=8cm]{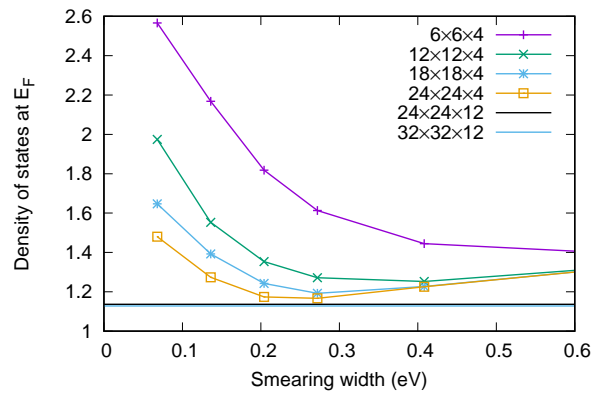}
\caption{Electronic density of states at the Fermi level with respect to the smearing width used in the integration, for various $k$-grids. The top four curves were obtained using the smearing integration of the density of states ($\tilde{N} (E_F)$), while the bottom two lines were obtained using tetrahedron integration (exact $N(E_F)$).}
\label{DOS}
\end{figure}

The magnitude of the error is well quantified by examining the electronic density of states $N(E) = \sum_{n\textbf{k}} \delta(E-\epsilon_{n\textbf{k}})$. We define $\tilde{N}$ as the density of states calculated using this smearing scheme. Here and hereafter, the functions and values with tilde denote the ones calculated with the smearing scheme. Equation \ref{a2F} indicates that $\tilde{\alpha}^2 F$ scales with $\tilde{N} (\tilde{E}_F)^2$ in the limit where the electron-phonon coupling coefficients $g$ are independent of $n$, $n'$, $\textbf{k}$ and $\textbf{q}$ \footnote{In the Phonon package of {\sc quantum espresso}, the prefactor $1/N(E_F)$ in equation \ref{a2F} is replaced by $1/\tilde{N} (E_F)$, hence in this case we only obtain a $\tilde{N} (E_F)$ dependence when summing the delta functions.}. To estimate the impact of the above quantitative errors, we examined the dependence of the electronic density of states at the Fermi energy $\tilde{N} (\tilde{E}_F)$ with respect to the smearing width and $k$-point grid (Figure \ref{DOS}). We obtained a well-converged value for $N(E_F)$ using a very dense $k$-point grid and the tetrahedron integration method, which we refer to as the ``exact" $N(E_F)$ hereafter. We found that $\tilde{N} (\tilde{E}_F)$ differs from the exact value by approximately 20\% at the Gaussian width 0.5 eV, which is within a standard range of the width. Moreover, we found slowly convergent behaviour for widths lower than 0.1 eV. Notably, the values of $\tilde{N} (\tilde{E}_F)$ are systematically larger than the exact one, indicating that $\tilde{\alpha}^2 F$ can suffer from systematic overestimation in this material.

\begin{figure}
\centering
\includegraphics[width=8cm]{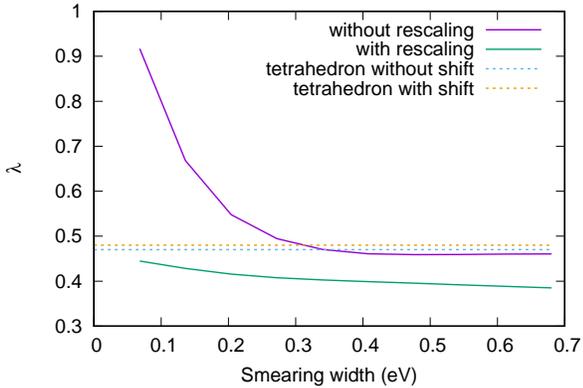}
\caption{Electron-phonon coupling strength $\lambda$ with respect to the smearing width used when integrating $\alpha^2F$. The top curve was obtained using the standard smearing method ($\tilde{\alpha}^2 F$) with a $\textbf{k}$-grid of $12 \times 12 \times 6$ points and a $\textbf{q}$-grid of $4 \times 4 \times 2$ without shift. The bottom curve was obtained using the rescaling method which mitigates the influence of errors in evaluating the electronic density of states ($\alpha^2 F_\text{resc}$). For reference, the values of $\lambda$ obtained using the tetrahedron method and two $\textbf{q}$-grids of $4 \times 4 \times 2$ points are also plotted.}
\label{lambda-figure}
\end{figure}

In order to mitigate the above-mentioned overestimation we modified the $\alpha^2 F$ formula in practice, following recent technical developments \cite{Koretsune2016}, and obtained the rescaled Eliashberg function \footnote{In the case of {\sc quantum espresso}, because the program uses the prefactor $1/\tilde{N} (E_F)$ in equation \ref{a2F}, we need to use the following formula:
\unexpanded{\begin{align}
\alpha^2 F_\text{resc} (\omega) = &\frac{\sum_{\textbf{q}} \sum_{n n' \textbf{k}} \delta (\epsilon_{n\textbf{k}+\textbf{q}} - E_F) \delta (\epsilon_{n'\textbf{k}} - E_F)}{\sum_{\textbf{q}} \sum_{n n' \textbf{k}} \tilde{\delta} (\epsilon_{n\textbf{k}+\textbf{q}} - \tilde{E}_F) \tilde{\delta} (\epsilon_{n'\textbf{k}} - \tilde{E}_F)} \nonumber \\ &\times \frac{\tilde{N} (\tilde{E}_F)}{N(E_F)} \tilde{\alpha^2 F} (\omega)	
\end{align}}}:
\begin{align}
\alpha^2 F_\text{resc} (\omega) \equiv &\frac{\sum_{\textbf{q}} \sum_{n n' \textbf{k}} \delta (\epsilon_{n\textbf{k}+\textbf{q}} - E_F) \delta (\epsilon_{n'\textbf{k}} - E_F)}{\sum_{\textbf{q}} \sum_{n n' \textbf{k}} \tilde{\delta} (\epsilon_{n\textbf{k}+\textbf{q}} - \tilde{E}_F) \tilde{\delta} (\epsilon_{n'\textbf{k}} - \tilde{E}_F)} \nonumber \\ &\times \tilde{\alpha^2 F} (\omega)
\label{a2F-rescaling}
\end{align}
The numerator is replaced to the exact $N(E_F)^2$ since it corresponds to the dense $\textbf{k}$ and $\textbf{q}$-points limit. Note that regardless of the k-point density, this modified formula gives us the exact value in the limit where the electron-phonon coupling coefficients $g$ are independent of $n$, $n'$, $\textbf{k}$ and $\textbf{q}$. With this formula, as shown in figure \ref{lambda-figure}, we obtained values of $\lambda$ almost independent of the smearing width.

We confirmed these results by calculating $\alpha^2 F$ with the optimized tetrahedron method \cite{Kawamura2014} (Figure \ref{a2F-figure}). We used four different $\textbf{q}$-point meshes, two with $3 \times 3 \times 3$ points (and a $18 \times 18 \times 6$ $\textbf{k}$-point mesh) and two with $4 \times 4 \times 2$ points (and a $24 \times 24 \times 4$ $\textbf{k}$-point mesh), one shifted along the $z$ axis and the other along all the axes \footnote{The unshifted $\textbf{q}$-mesh cannot be employed for the tetrahedron integration since it suffers from numerical divergence at $\textbf{q} = 0$. Note that this is an artefact due to the discretisation and can be avoided if the $\textbf{q}$-point integral around $\textbf{q} = 0$ is calculated analytically} \footnote{We have also employed the rescaling formula (Equation \eqref{a2F-rescaling}) for $\alpha^2 F_{tetra}$ to accelerate the $\textbf{k}$-point convergence. In this case, the denominator of the prefactor is calculated with the $\textbf{k}$ and $\textbf{q}$-point mesh employed for the integration in equation \eqref{a2F}.}. We name $\alpha^2 F_{\text{tetra}}$ the Eliashberg function obtained using the all-shifted $4 \times 4 \times 2$ $\textbf{q}$-point mesh. We obtained $\lambda = 0.39$ and $\lambda = 0.58$ for the first sampling, with $z$-shifted and all-shifted meshes, respectively, and $\lambda = 0.48$ and $\lambda = 0.47$ for the second sampling. The values obtained with $3 \times 3 \times 3$ points differ clearly, indicating a lack of convergence, while the ones obtained with $4 \times 4 \times 2$ points are both very close and between the two previous ones. This indicates a strong convergence. Moreover, the values obtained with the $4 \times 4 \times 2$ $\textbf{q}$-point mesh match closely the values obtained by integrating $\alpha^2 F_\text{resc}$ in the limit of small smearing (Figure \ref{lambda-figure}), which is the exact limit. The small deviation between the values of $\lambda$ indicates that the converged $\lambda$ is about or below 0.5.

The results obtained for $\omega_{ln}$ are less conclusive: indeed the values obtained for the two $3 \times 3 \times 3$ $\textbf{q}$-point grids are 116 K and 153 K, while the ones obtained using the two $4 \times 4 \times 2$ $\textbf{q}$-point grids are 98 K and 143 K. Let us note that $T_c$ is affected exponentially by $\lambda$ whereas it is affected only linearly by $\omega_{ln}$ (Equation \eqref{McMillan-Allen-Dynes}), which means that the convergence of $\lambda$ gives us the order of magnitude of $T_c$. In the following subsection, we show that this order of magnitude is clearly different from the one observed experimentally.

\subsection{Superconducting transition temperature}

We calculated the superconducting transition temperature ($T_c$) of LaO$_{0.5}$F$_{0.5}$BiS$_2$ using density functional theory for superconductors \footnote{We sampled the Brillouin zone and the different bands using a weighted random set of points \cite{Marques2005a}. The standard error on the value of the gap was found to be 18.3\%.}. We ran a ``static" calculation which only uses the zero frequency dielectric function, and a ``dynamical" calculation which uses the full frequency-dependent dielectric function, both calculated using a $4 \times 4 \times 2$ $\textbf{k}$-point grid. We used $\alpha^2 F_\text{tetra}$ in both calculations. The results are summarised in table \ref{table}.

\begin{table}
\begin{tabular}{c|cc}
& static & dynamical\\
\hline
$T_c$ & 0.2 K & 0.4 K\\
$\mu^*$ & 0.18 & 0.16
\end{tabular}
\caption{Superconducting transition temperatures and effective Coulomb repulsion factors in LaO$_{0.5}$F$_{0.5}$BiS$_2$ calculated using density functional theory for superconductors.}
\label{table}
\end{table}

It is interesting to compare this result to the output of the McMillan-Allen-Dynes formula \cite{McMillan1968}. We calculated $T_c$ with this formula using the parameters corresponding to $\alpha^2 F_{\text{tetra}}$ for a large range of values of $\mu^*$. We then can obtain the $\mu^*$ corresponding to the value of $T_c$ calculated using density functional theory for superconductors. We obtain values of $\mu^*$ larger than in most conventional superconductors \cite{Carbotte1990, Akashi2012} (Table \ref{table}). This is reminiscent of other ab-initio studies of superconductors \cite{Akashi2012}.

The value of $T_c$ calculated here are very distant from all the values of $T_c$ measured in this compound, which are between 2.5 and 11.5 K \cite{Mizuguchi2014a}. We therefore conclude that conventional phonon-coupling superconductivity cannot explain the experimental data on LaO$_{0.5}$F$_{0.5}$BiS$_2$.

\section{Conclusion}

We calculated phonon frequencies in the original unit cell of LaO$_{0.5}$F$_{0.5}$BiS$_2$, using two different approximations and spin-orbit coupling, and found a structural instability close to the M point of the Brillouin zone. This confirms previous findings \cite{Wan2013, Yildirim2013}.

The electron-phonon coupling was calculated in a relaxed $\sqrt{2} \times \sqrt{2}$ structure using the standard smearing integration scheme and obtained values for $\lambda$ with a strong smearing width dependence (Figure \ref{lambda-figure}). These results match previous calculations well \cite{Yildirim2013}.

We then used a cutting-edge rescaling technique \cite{Koretsune2016} to mitigate the influence of errors in the calculation of the density of states, which resulted in a much smaller dependence of $\lambda$ on smearing width (Figure \ref{lambda-figure}). The obtained $\lambda$ matches closely results obtained with a novel tetrahedron method \cite{Kawamura2014}. We conclude that $\lambda < 0.5$.

Finally, using the Eliashberg function calculated using the tetrahedron method, we calculated $T_c$ using density functional theory for superconductors \cite{Luders2005,Marques2005a}. We obtained $T_c = 0.4$ K. This corresponds to a pair-breaking electron-electron repulsion $\mu^* = 0.16$, which is larger than conventional values. We thus conclude that conventional phonon-mediated pairing cannot explain the high $T_c$s observed in this material. This is an important step toward the clarification of the origin of superconductivity in BiS$_2$ superconductors.

\section*{Acknowledgments}

We would like to thank K. Kuroki and H. Ikeda for interesting discussions. We acknowledge support from EPSRC, Corpus Christi College, and JST, PRESTO and JSPS KAKENHI grants number JP15H03696, JP16H00924, JP16H06345 and JP15K20940. This work was supported by MEXT Element Strategy Initiative to Form Core Research Center in Japan. Part of the calculations were performed at the Supercomputer Center at the Institute for Solid State Physics in the University of Tokyo.

\bibliographystyle{apsrev4-1}
\bibliography{library}

\begin{thebibliography}{65}%
\makeatletter
\providecommand \@ifxundefined [1]{%
 \@ifx{#1\undefined}
}%
\providecommand \@ifnum [1]{%
 \ifnum #1\expandafter \@firstoftwo
 \else \expandafter \@secondoftwo
 \fi
}%
\providecommand \@ifx [1]{%
 \ifx #1\expandafter \@firstoftwo
 \else \expandafter \@secondoftwo
 \fi
}%
\providecommand \natexlab [1]{#1}%
\providecommand \enquote  [1]{``#1''}%
\providecommand \bibnamefont  [1]{#1}%
\providecommand \bibfnamefont [1]{#1}%
\providecommand \citenamefont [1]{#1}%
\providecommand \href@noop [0]{\@secondoftwo}%
\providecommand \href [0]{\begingroup \@sanitize@url \@href}%
\providecommand \@href[1]{\@@startlink{#1}\@@href}%
\providecommand \@@href[1]{\endgroup#1\@@endlink}%
\providecommand \@sanitize@url [0]{\catcode `\\12\catcode `\$12\catcode
  `\&12\catcode `\#12\catcode `\^12\catcode `\_12\catcode `\%12\relax}%
\providecommand \@@startlink[1]{}%
\providecommand \@@endlink[0]{}%
\providecommand \url  [0]{\begingroup\@sanitize@url \@url }%
\providecommand \@url [1]{\endgroup\@href {#1}{\urlprefix }}%
\providecommand \urlprefix  [0]{URL }%
\providecommand \Eprint [0]{\href }%
\providecommand \doibase [0]{http://dx.doi.org/}%
\providecommand \selectlanguage [0]{\@gobble}%
\providecommand \bibinfo  [0]{\@secondoftwo}%
\providecommand \bibfield  [0]{\@secondoftwo}%
\providecommand \translation [1]{[#1]}%
\providecommand \BibitemOpen [0]{}%
\providecommand \bibitemStop [0]{}%
\providecommand \bibitemNoStop [0]{.\EOS\space}%
\providecommand \EOS [0]{\spacefactor3000\relax}%
\providecommand \BibitemShut  [1]{\csname bibitem#1\endcsname}%
\let\auto@bib@innerbib\@empty
\bibitem [{\citenamefont {Mizuguchi}\ \emph
  {et~al.}(2012{\natexlab{a}})\citenamefont {Mizuguchi}, \citenamefont
  {Fujihisa}, \citenamefont {Gotoh}, \citenamefont {Suzuki}, \citenamefont
  {Usui}, \citenamefont {Kuroki}, \citenamefont {Demura}, \citenamefont
  {Takano}, \citenamefont {Izawa},\ and\ \citenamefont
  {Miura}}]{Mizuguchi2012}%
  \BibitemOpen
  \bibfield  {author} {\bibinfo {author} {\bibfnamefont {Y.}~\bibnamefont
  {Mizuguchi}}, \bibinfo {author} {\bibfnamefont {H.}~\bibnamefont {Fujihisa}},
  \bibinfo {author} {\bibfnamefont {Y.}~\bibnamefont {Gotoh}}, \bibinfo
  {author} {\bibfnamefont {K.}~\bibnamefont {Suzuki}}, \bibinfo {author}
  {\bibfnamefont {H.}~\bibnamefont {Usui}}, \bibinfo {author} {\bibfnamefont
  {K.}~\bibnamefont {Kuroki}}, \bibinfo {author} {\bibfnamefont
  {S.}~\bibnamefont {Demura}}, \bibinfo {author} {\bibfnamefont
  {Y.}~\bibnamefont {Takano}}, \bibinfo {author} {\bibfnamefont
  {H.}~\bibnamefont {Izawa}}, \ and\ \bibinfo {author} {\bibfnamefont
  {O.}~\bibnamefont {Miura}},\ }\href {\doibase 10.1103/PhysRevB.86.220510}
  {\bibfield  {journal} {\bibinfo  {journal} {Phys. Rev. B}\ }\textbf {\bibinfo
  {volume} {86}},\ \bibinfo {pages} {220510} (\bibinfo {year}
  {2012}{\natexlab{a}})}\BibitemShut {NoStop}%
\bibitem [{\citenamefont {Mizuguchi}\ \emph
  {et~al.}(2012{\natexlab{b}})\citenamefont {Mizuguchi}, \citenamefont
  {Demura}, \citenamefont {Deguchi}, \citenamefont {Takano}, \citenamefont
  {Fujihisa}, \citenamefont {Gotoh}, \citenamefont {Izawa},\ and\ \citenamefont
  {Miura}}]{Mizuguchi2012a}%
  \BibitemOpen
  \bibfield  {author} {\bibinfo {author} {\bibfnamefont {Y.}~\bibnamefont
  {Mizuguchi}}, \bibinfo {author} {\bibfnamefont {S.}~\bibnamefont {Demura}},
  \bibinfo {author} {\bibfnamefont {K.}~\bibnamefont {Deguchi}}, \bibinfo
  {author} {\bibfnamefont {Y.}~\bibnamefont {Takano}}, \bibinfo {author}
  {\bibfnamefont {H.}~\bibnamefont {Fujihisa}}, \bibinfo {author}
  {\bibfnamefont {Y.}~\bibnamefont {Gotoh}}, \bibinfo {author} {\bibfnamefont
  {H.}~\bibnamefont {Izawa}}, \ and\ \bibinfo {author} {\bibfnamefont
  {O.}~\bibnamefont {Miura}},\ }\href {\doibase 10.7566/JPSJ.83.063704}
  {\bibfield  {journal} {\bibinfo  {journal} {J. Phys. Soc. Japan}\ }\textbf
  {\bibinfo {volume} {81}},\ \bibinfo {pages} {114725} (\bibinfo {year}
  {2012}{\natexlab{b}})},\ \BibitemShut {NoStop}%
\bibitem [{\citenamefont {Awana}\ \emph {et~al.}(2013)\citenamefont {Awana},
  \citenamefont {Kumar}, \citenamefont {Jha}, \citenamefont {Singh},
  \citenamefont {Pal}, \citenamefont {Shruti}, \citenamefont {Saha},\ and\
  \citenamefont {Patnaik}}]{Awana2013}%
  \BibitemOpen
  \bibfield  {author} {\bibinfo {author} {\bibfnamefont {V.~P.~S.}\
  \bibnamefont {Awana}}, \bibinfo {author} {\bibfnamefont {A.}~\bibnamefont
  {Kumar}}, \bibinfo {author} {\bibfnamefont {R.}~\bibnamefont {Jha}}, \bibinfo
  {author} {\bibfnamefont {S.~K.}\ \bibnamefont {Singh}}, \bibinfo {author}
  {\bibfnamefont {A.}~\bibnamefont {Pal}}, \bibinfo {author} {\bibnamefont
  {Shruti}}, \bibinfo {author} {\bibfnamefont {J.}~\bibnamefont {Saha}}, \ and\
  \bibinfo {author} {\bibfnamefont {S.}~\bibnamefont {Patnaik}},\ }\href
  {\doibase 10.1016/j.ssc.2012.11.021} {\bibfield  {journal} {\bibinfo
  {journal} {Solid State Commun.}\ }\textbf {\bibinfo {volume} {157}},\
  \bibinfo {pages} {21} (\bibinfo {year} {2013})}\BibitemShut {NoStop}%
\bibitem [{\citenamefont {Usui}\ \emph {et~al.}(2012)\citenamefont {Usui},
  \citenamefont {Suzuki},\ and\ \citenamefont {Kuroki}}]{Usui2012}%
  \BibitemOpen
  \bibfield  {author} {\bibinfo {author} {\bibfnamefont {H.}~\bibnamefont
  {Usui}}, \bibinfo {author} {\bibfnamefont {K.}~\bibnamefont {Suzuki}}, \ and\
  \bibinfo {author} {\bibfnamefont {K.}~\bibnamefont {Kuroki}},\ }\href
  {\doibase 10.1103/PhysRevB.86.220501} {\bibfield  {journal} {\bibinfo
  {journal} {Phys. Rev. B}\ }\textbf {\bibinfo {volume} {86}},\ \bibinfo
  {pages} {220501} (\bibinfo {year} {2012})}\BibitemShut {NoStop}%
\bibitem [{\citenamefont {Usui}\ and\ \citenamefont {Kuroki}(2015)}]{Usui2015}%
  \BibitemOpen
  \bibfield  {author} {\bibinfo {author} {\bibfnamefont {H.}~\bibnamefont
  {Usui}}\ and\ \bibinfo {author} {\bibfnamefont {K.}~\bibnamefont {Kuroki}},\
  }\href {\doibase 10.1515/nsm-2015-0005} {\bibfield  {journal} {\bibinfo
  {journal} {Nov. Supercond. Mater.}\ }\textbf {\bibinfo {volume} {1}},\
  \bibinfo {pages} {27} (\bibinfo {year} {2015})},\  \BibitemShut {NoStop}%
\bibitem [{\citenamefont {Morice}\ \emph {et~al.}(2015)\citenamefont {Morice},
  \citenamefont {Artacho}, \citenamefont {Dutton}, \citenamefont {Molnar},
  \citenamefont {Kim},\ and\ \citenamefont {Saxena}}]{Morice2015}%
  \BibitemOpen
  \bibfield  {author} {\bibinfo {author} {\bibfnamefont {C.}~\bibnamefont
  {Morice}}, \bibinfo {author} {\bibfnamefont {E.}~\bibnamefont {Artacho}},
  \bibinfo {author} {\bibfnamefont {S.~E.}\ \bibnamefont {Dutton}}, \bibinfo
  {author} {\bibfnamefont {D.}~\bibnamefont {Molnar}}, \bibinfo {author}
  {\bibfnamefont {H.-J.}\ \bibnamefont {Kim}}, \ and\ \bibinfo {author}
  {\bibfnamefont {S.~S.}\ \bibnamefont {Saxena}},\ }\href
  {http://stacks.iop.org/0953-8984/27/i=13/a=135501} {\bibfield  {journal}
  {\bibinfo  {journal} {J. Phys. Condens. Matter}\ }\textbf {\bibinfo {volume}
  {27}},\ \bibinfo {pages} {135501} (\bibinfo {year} {2015})}\BibitemShut
  {NoStop}%
\bibitem [{\citenamefont {Xing}\ \emph {et~al.}(2012)\citenamefont {Xing},
  \citenamefont {Li}, \citenamefont {Ding}, \citenamefont {Yang},\ and\
  \citenamefont {Wen}}]{Xing2012}%
  \BibitemOpen
  \bibfield  {author} {\bibinfo {author} {\bibfnamefont {J.}~\bibnamefont
  {Xing}}, \bibinfo {author} {\bibfnamefont {S.}~\bibnamefont {Li}}, \bibinfo
  {author} {\bibfnamefont {X.}~\bibnamefont {Ding}}, \bibinfo {author}
  {\bibfnamefont {H.}~\bibnamefont {Yang}}, \ and\ \bibinfo {author}
  {\bibfnamefont {H.-H.}\ \bibnamefont {Wen}},\ }\href {\doibase
  10.1103/PhysRevB.86.214518} {\bibfield  {journal} {\bibinfo  {journal} {Phys.
  Rev. B}\ }\textbf {\bibinfo {volume} {86}},\ \bibinfo {pages} {214518}
  (\bibinfo {year} {2012})}\BibitemShut {NoStop}%
\bibitem [{\citenamefont {Jha}\ and\ \citenamefont {Awana}(2013)}]{Jha2013c}%
  \BibitemOpen
  \bibfield  {author} {\bibinfo {author} {\bibfnamefont {R.}~\bibnamefont
  {Jha}}\ and\ \bibinfo {author} {\bibfnamefont {V.~P.~S.}\ \bibnamefont
  {Awana}},\ }\href {\doibase 10.1007/s10948-013-2404-0} {\bibfield  {journal}
  {\bibinfo  {journal} {J. Supercond. Nov. Magn.}\ }\textbf {\bibinfo {volume}
  {27}},\ \bibinfo {pages} {1} (\bibinfo {year} {2013})},\  \BibitemShut {NoStop}%
\bibitem [{\citenamefont {Demura}\ \emph {et~al.}(2013)\citenamefont {Demura},
  \citenamefont {Mizuguchi}, \citenamefont {Deguchi}, \citenamefont {Okazaki},
  \citenamefont {Hara}, \citenamefont {Watanabe}, \citenamefont {Denholme},
  \citenamefont {Fujioka}, \citenamefont {Ozaki}, \citenamefont {Fujuhisa},
  \citenamefont {Gotoh}, \citenamefont {Miura}, \citenamefont {Yamaguchi},
  \citenamefont {Takeya},\ and\ \citenamefont {Takano}}]{Demura2013}%
  \BibitemOpen
  \bibfield  {author} {\bibinfo {author} {\bibfnamefont {S.}~\bibnamefont
  {Demura}}, \bibinfo {author} {\bibfnamefont {Y.}~\bibnamefont {Mizuguchi}},
  \bibinfo {author} {\bibfnamefont {K.}~\bibnamefont {Deguchi}}, \bibinfo
  {author} {\bibfnamefont {H.}~\bibnamefont {Okazaki}}, \bibinfo {author}
  {\bibfnamefont {H.}~\bibnamefont {Hara}}, \bibinfo {author} {\bibfnamefont
  {T.}~\bibnamefont {Watanabe}}, \bibinfo {author} {\bibfnamefont {S.~J.}\
  \bibnamefont {Denholme}}, \bibinfo {author} {\bibfnamefont {M.}~\bibnamefont
  {Fujioka}}, \bibinfo {author} {\bibfnamefont {T.}~\bibnamefont {Ozaki}},
  \bibinfo {author} {\bibfnamefont {H.}~\bibnamefont {Fujuhisa}}, \bibinfo
  {author} {\bibfnamefont {Y.}~\bibnamefont {Gotoh}}, \bibinfo {author}
  {\bibfnamefont {O.}~\bibnamefont {Miura}}, \bibinfo {author} {\bibfnamefont
  {T.}~\bibnamefont {Yamaguchi}}, \bibinfo {author} {\bibfnamefont
  {H.}~\bibnamefont {Takeya}}, \ and\ \bibinfo {author} {\bibfnamefont
  {Y.}~\bibnamefont {Takano}},\ }\href {\doibase 10.1088/0953-2048/29/1/015007}
  {\bibfield  {journal} {\bibinfo  {journal} {J. Phys. Soc. Japan}\ }\textbf
  {\bibinfo {volume} {82}},\ \bibinfo {pages} {033708} (\bibinfo {year}
  {2013})}\BibitemShut {NoStop}%
\bibitem [{\citenamefont {Jha}\ \emph {et~al.}(2013)\citenamefont {Jha},
  \citenamefont {Kumar}, \citenamefont {{Kumar Singh}},\ and\ \citenamefont
  {Awana}}]{Jha2013a}%
  \BibitemOpen
  \bibfield  {author} {\bibinfo {author} {\bibfnamefont {R.}~\bibnamefont
  {Jha}}, \bibinfo {author} {\bibfnamefont {A.}~\bibnamefont {Kumar}}, \bibinfo
  {author} {\bibfnamefont {S.}~\bibnamefont {{Kumar Singh}}}, \ and\ \bibinfo
  {author} {\bibfnamefont {V.~P.~S.}\ \bibnamefont {Awana}},\ }\href {\doibase
  10.1007/s10948-012-2097-9} {\bibfield  {journal} {\bibinfo  {journal} {J.
  Supercond. Nov. Magn.}\ }\textbf {\bibinfo {volume} {26}},\ \bibinfo {pages}
  {499} (\bibinfo {year} {2013})}\BibitemShut {NoStop}%
\bibitem [{\citenamefont {Zhai}\ \emph {et~al.}(2014)\citenamefont {Zhai},
  \citenamefont {Tang}, \citenamefont {Jiang}, \citenamefont {Xu},
  \citenamefont {Zhang}, \citenamefont {Zhang}, \citenamefont {Bao},
  \citenamefont {Sun}, \citenamefont {Jiao}, \citenamefont {Nowik},
  \citenamefont {Felner}, \citenamefont {Li}, \citenamefont {Xu}, \citenamefont
  {Tao}, \citenamefont {Feng}, \citenamefont {Xu},\ and\ \citenamefont
  {Cao}}]{Zhai2014}%
  \BibitemOpen
  \bibfield  {author} {\bibinfo {author} {\bibfnamefont {H.-F.}\ \bibnamefont
  {Zhai}}, \bibinfo {author} {\bibfnamefont {Z.-T.}\ \bibnamefont {Tang}},
  \bibinfo {author} {\bibfnamefont {H.}~\bibnamefont {Jiang}}, \bibinfo
  {author} {\bibfnamefont {K.}~\bibnamefont {Xu}}, \bibinfo {author}
  {\bibfnamefont {K.}~\bibnamefont {Zhang}}, \bibinfo {author} {\bibfnamefont
  {P.}~\bibnamefont {Zhang}}, \bibinfo {author} {\bibfnamefont {J.-K.}\
  \bibnamefont {Bao}}, \bibinfo {author} {\bibfnamefont {Y.-L.}\ \bibnamefont
  {Sun}}, \bibinfo {author} {\bibfnamefont {W.-H.}\ \bibnamefont {Jiao}},
  \bibinfo {author} {\bibfnamefont {I.}~\bibnamefont {Nowik}}, \bibinfo
  {author} {\bibfnamefont {I.}~\bibnamefont {Felner}}, \bibinfo {author}
  {\bibfnamefont {Y.-K.}\ \bibnamefont {Li}}, \bibinfo {author} {\bibfnamefont
  {X.-F.}\ \bibnamefont {Xu}}, \bibinfo {author} {\bibfnamefont
  {Q.}~\bibnamefont {Tao}}, \bibinfo {author} {\bibfnamefont {C.-M.}\
  \bibnamefont {Feng}}, \bibinfo {author} {\bibfnamefont {Z.-A.}\ \bibnamefont
  {Xu}}, \ and\ \bibinfo {author} {\bibfnamefont {G.-H.}\ \bibnamefont {Cao}},\
  }\href {\doibase 10.1103/PhysRevB.90.064518} {\bibfield  {journal} {\bibinfo
  {journal} {Phys. Rev. B}\ }\textbf {\bibinfo {volume} {90}},\ \bibinfo
  {pages} {64518} (\bibinfo {year} {2014})}\BibitemShut {NoStop}%
\bibitem [{\citenamefont {Yazici}\ \emph {et~al.}(2013)\citenamefont {Yazici},
  \citenamefont {Huang}, \citenamefont {White}, \citenamefont {Chang},
  \citenamefont {Friedman},\ and\ \citenamefont {Maple}}]{Yazici2013b}%
  \BibitemOpen
  \bibfield  {author} {\bibinfo {author} {\bibfnamefont {D.}~\bibnamefont
  {Yazici}}, \bibinfo {author} {\bibfnamefont {K.}~\bibnamefont {Huang}},
  \bibinfo {author} {\bibfnamefont {B.~D.}\ \bibnamefont {White}}, \bibinfo
  {author} {\bibfnamefont {A.~H.}\ \bibnamefont {Chang}}, \bibinfo {author}
  {\bibfnamefont {A.~J.}\ \bibnamefont {Friedman}}, \ and\ \bibinfo {author}
  {\bibfnamefont {M.~B.}\ \bibnamefont {Maple}},\ }\href {\doibase
  10.1080/14786435.2012.724185} {\bibfield  {journal} {\bibinfo  {journal}
  {Philos. Mag.}\ }\textbf {\bibinfo {volume} {93}},\ \bibinfo {pages} {673}
  (\bibinfo {year} {2013})}\BibitemShut {NoStop}%
\bibitem [{\citenamefont {Wan}\ \emph {et~al.}(2013)\citenamefont {Wan},
  \citenamefont {Ding}, \citenamefont {Savrasov},\ and\ \citenamefont
  {Duan}}]{Wan2013}%
  \BibitemOpen
  \bibfield  {author} {\bibinfo {author} {\bibfnamefont {X.}~\bibnamefont
  {Wan}}, \bibinfo {author} {\bibfnamefont {H.-C.}\ \bibnamefont {Ding}},
  \bibinfo {author} {\bibfnamefont {S.~Y.}\ \bibnamefont {Savrasov}}, \ and\
  \bibinfo {author} {\bibfnamefont {C.-G.}\ \bibnamefont {Duan}},\ }\href
  {\doibase 10.1103/PhysRevB.87.115124} {\bibfield  {journal} {\bibinfo
  {journal} {Phys. Rev. B}\ }\textbf {\bibinfo {volume} {87}},\ \bibinfo
  {pages} {115124} (\bibinfo {year} {2013})}\BibitemShut {NoStop}%
\bibitem [{\citenamefont {Yildirim}(2013)}]{Yildirim2013}%
  \BibitemOpen
  \bibfield  {author} {\bibinfo {author} {\bibfnamefont {T.}~\bibnamefont
  {Yildirim}},\ }\href {\doibase 10.1103/PhysRevB.87.020506} {\bibfield
  {journal} {\bibinfo  {journal} {Phys. Rev. B}\ }\textbf {\bibinfo {volume}
  {87}},\ \bibinfo {pages} {20506} (\bibinfo {year} {2013})}\BibitemShut
  {NoStop}%
\bibitem [{\citenamefont {Morice}\ \emph {et~al.}(2016)\citenamefont {Morice},
  \citenamefont {Artacho}, \citenamefont {Dutton}, \citenamefont {Kim},\ and\
  \citenamefont {Saxena}}]{Morice2016}%
  \BibitemOpen
  \bibfield  {author} {\bibinfo {author} {\bibfnamefont {C.}~\bibnamefont
  {Morice}}, \bibinfo {author} {\bibfnamefont {E.}~\bibnamefont {Artacho}},
  \bibinfo {author} {\bibfnamefont {S.~E.}\ \bibnamefont {Dutton}}, \bibinfo
  {author} {\bibfnamefont {H.-J.}\ \bibnamefont {Kim}}, \ and\ \bibinfo
  {author} {\bibfnamefont {S.~S.}\ \bibnamefont {Saxena}},\ }\href {\doibase
  10.1088/0953-8984/28/34/345504} {\bibfield  {journal} {\bibinfo  {journal}
  {J. Phys. Condens. Matter}\ }\textbf {\bibinfo {volume} {28}},\ \bibinfo
  {pages} {345504} (\bibinfo {year} {2016})},\  \BibitemShut {NoStop}%
\bibitem [{\citenamefont {Higashinaka}\ \emph {et~al.}(2015)\citenamefont
  {Higashinaka}, \citenamefont {Asano}, \citenamefont {Nakashima},
  \citenamefont {Fushiya}, \citenamefont {Mizuguchi}, \citenamefont {Miura},
  \citenamefont {Matsuda},\ and\ \citenamefont {Aoki}}]{Higashinaka2015}%
  \BibitemOpen
  \bibfield  {author} {\bibinfo {author} {\bibfnamefont {R.}~\bibnamefont
  {Higashinaka}}, \bibinfo {author} {\bibfnamefont {T.}~\bibnamefont {Asano}},
  \bibinfo {author} {\bibfnamefont {T.}~\bibnamefont {Nakashima}}, \bibinfo
  {author} {\bibfnamefont {K.}~\bibnamefont {Fushiya}}, \bibinfo {author}
  {\bibfnamefont {Y.}~\bibnamefont {Mizuguchi}}, \bibinfo {author}
  {\bibfnamefont {O.}~\bibnamefont {Miura}}, \bibinfo {author} {\bibfnamefont
  {T.~D.}\ \bibnamefont {Matsuda}}, \ and\ \bibinfo {author} {\bibfnamefont
  {Y.}~\bibnamefont {Aoki}},\ }\href {\doibase 10.7566/JPSJ.84.023702}
  {\bibfield  {journal} {\bibinfo  {journal} {J. Phys. Soc. Japan}\ }\textbf
  {\bibinfo {volume} {84}},\ \bibinfo {pages} {023702} (\bibinfo {year}
  {2015})},\  \BibitemShut {NoStop}%
\bibitem [{\citenamefont {Mizuguchi}\ \emph {et~al.}(2014)\citenamefont
  {Mizuguchi}, \citenamefont {Hiroi}, \citenamefont {Kajitani},\ and\
  \citenamefont {Takatsu}}]{Mizuguchi2014a}%
  \BibitemOpen
  \bibfield  {author} {\bibinfo {author} {\bibfnamefont {Y.}~\bibnamefont
  {Mizuguchi}}, \bibinfo {author} {\bibfnamefont {T.}~\bibnamefont {Hiroi}},
  \bibinfo {author} {\bibfnamefont {J.}~\bibnamefont {Kajitani}}, \ and\
  \bibinfo {author} {\bibfnamefont {H.}~\bibnamefont {Takatsu}},\ }\href@noop
  {} {\bibfield  {journal} {\bibinfo  {journal} {J. Phys. Soc. Japan}\ }\textbf
  {\bibinfo {volume} {83}},\ \bibinfo {pages} {053704} (\bibinfo {year}
  {2014})}\BibitemShut {NoStop}%
\bibitem [{\citenamefont {Zhang}\ \emph {et~al.}(2016)\citenamefont {Zhang},
  \citenamefont {Huang}, \citenamefont {Ding}, \citenamefont {MacLaughlin},
  \citenamefont {Bernal}, \citenamefont {Ho}, \citenamefont {Tan},
  \citenamefont {Liu}, \citenamefont {Yazici}, \citenamefont {Maple},\ and\
  \citenamefont {Shu}}]{Zhang2016}%
  \BibitemOpen
  \bibfield  {author} {\bibinfo {author} {\bibfnamefont {J.}~\bibnamefont
  {Zhang}}, \bibinfo {author} {\bibfnamefont {K.}~\bibnamefont {Huang}},
  \bibinfo {author} {\bibfnamefont {Z.~F.}\ \bibnamefont {Ding}}, \bibinfo
  {author} {\bibfnamefont {D.~E.}\ \bibnamefont {MacLaughlin}}, \bibinfo
  {author} {\bibfnamefont {O.~O.}\ \bibnamefont {Bernal}}, \bibinfo {author}
  {\bibfnamefont {P.~C.}\ \bibnamefont {Ho}}, \bibinfo {author} {\bibfnamefont
  {C.}~\bibnamefont {Tan}}, \bibinfo {author} {\bibfnamefont {X.}~\bibnamefont
  {Liu}}, \bibinfo {author} {\bibfnamefont {D.}~\bibnamefont {Yazici}},
  \bibinfo {author} {\bibfnamefont {M.~B.}\ \bibnamefont {Maple}}, \ and\
  \bibinfo {author} {\bibfnamefont {L.}~\bibnamefont {Shu}},\ }\href
  {http://arxiv.org/abs/1611.04726} {\ \textbf {\bibinfo {volume} {2}},\
  \bibinfo {pages} {1} (\bibinfo {year} {2016})},\  \BibitemShut {NoStop}%
\bibitem [{\citenamefont {Deguchi}\ \emph {et~al.}(2013)\citenamefont
  {Deguchi}, \citenamefont {Mizuguchi}, \citenamefont {Demura}, \citenamefont
  {Hara}, \citenamefont {Watanabe}, \citenamefont {Denholme}, \citenamefont
  {Fujioka}, \citenamefont {Okazaki}, \citenamefont {Ozaki}, \citenamefont
  {Takeya}, \citenamefont {Yamaguchi}, \citenamefont {Miura},\ and\
  \citenamefont {Takano}}]{Deguchi2013}%
  \BibitemOpen
  \bibfield  {author} {\bibinfo {author} {\bibfnamefont {K.}~\bibnamefont
  {Deguchi}}, \bibinfo {author} {\bibfnamefont {Y.}~\bibnamefont {Mizuguchi}},
  \bibinfo {author} {\bibfnamefont {S.}~\bibnamefont {Demura}}, \bibinfo
  {author} {\bibfnamefont {H.}~\bibnamefont {Hara}}, \bibinfo {author}
  {\bibfnamefont {T.}~\bibnamefont {Watanabe}}, \bibinfo {author}
  {\bibfnamefont {S.~J.}\ \bibnamefont {Denholme}}, \bibinfo {author}
  {\bibfnamefont {M.}~\bibnamefont {Fujioka}}, \bibinfo {author} {\bibfnamefont
  {H.}~\bibnamefont {Okazaki}}, \bibinfo {author} {\bibfnamefont
  {T.}~\bibnamefont {Ozaki}}, \bibinfo {author} {\bibfnamefont
  {H.}~\bibnamefont {Takeya}}, \bibinfo {author} {\bibfnamefont
  {T.}~\bibnamefont {Yamaguchi}}, \bibinfo {author} {\bibfnamefont
  {O.}~\bibnamefont {Miura}}, \ and\ \bibinfo {author} {\bibfnamefont
  {Y.}~\bibnamefont {Takano}},\ }\href
  {http://stacks.iop.org/0295-5075/101/i=1/a=17004} {\bibfield  {journal}
  {\bibinfo  {journal} {Europhys. Lett.}\ }\textbf {\bibinfo {volume} {101}},\
  \bibinfo {pages} {17004} (\bibinfo {year} {2013})}\BibitemShut {NoStop}%
\bibitem [{\citenamefont {Kajitani}\ \emph {et~al.}(2014)\citenamefont
  {Kajitani}, \citenamefont {Deguchi}, \citenamefont {Omachi}, \citenamefont
  {Hiroi}, \citenamefont {Takano}, \citenamefont {Takatsu}, \citenamefont
  {Kadowaki}, \citenamefont {Miura},\ and\ \citenamefont
  {Mizuguchi}}]{Kajitani2014}%
  \BibitemOpen
  \bibfield  {author} {\bibinfo {author} {\bibfnamefont {J.}~\bibnamefont
  {Kajitani}}, \bibinfo {author} {\bibfnamefont {K.}~\bibnamefont {Deguchi}},
  \bibinfo {author} {\bibfnamefont {A.}~\bibnamefont {Omachi}}, \bibinfo
  {author} {\bibfnamefont {T.}~\bibnamefont {Hiroi}}, \bibinfo {author}
  {\bibfnamefont {Y.}~\bibnamefont {Takano}}, \bibinfo {author} {\bibfnamefont
  {H.}~\bibnamefont {Takatsu}}, \bibinfo {author} {\bibfnamefont
  {H.}~\bibnamefont {Kadowaki}}, \bibinfo {author} {\bibfnamefont
  {O.}~\bibnamefont {Miura}}, \ and\ \bibinfo {author} {\bibfnamefont
  {Y.}~\bibnamefont {Mizuguchi}},\ }\href {\doibase 10.1016/j.ssc.2013.11.027}
  {\bibfield  {journal} {\bibinfo  {journal} {Solid State Commun.}\ }\textbf
  {\bibinfo {volume} {181}},\ \bibinfo {pages} {1} (\bibinfo {year} {2014})},\
   \BibitemShut
  {NoStop}%
\bibitem [{\citenamefont {Pallecchi}\ \emph {et~al.}(2014)\citenamefont
  {Pallecchi}, \citenamefont {Lamura}, \citenamefont {Putti}, \citenamefont
  {Kajitani}, \citenamefont {Mizuguchi}, \citenamefont {Miura}, \citenamefont
  {Demura}, \citenamefont {Deguchi},\ and\ \citenamefont
  {Takano}}]{Pallecchi2014}%
  \BibitemOpen
  \bibfield  {author} {\bibinfo {author} {\bibfnamefont {I.}~\bibnamefont
  {Pallecchi}}, \bibinfo {author} {\bibfnamefont {G.}~\bibnamefont {Lamura}},
  \bibinfo {author} {\bibfnamefont {M.}~\bibnamefont {Putti}}, \bibinfo
  {author} {\bibfnamefont {J.}~\bibnamefont {Kajitani}}, \bibinfo {author}
  {\bibfnamefont {Y.}~\bibnamefont {Mizuguchi}}, \bibinfo {author}
  {\bibfnamefont {O.}~\bibnamefont {Miura}}, \bibinfo {author} {\bibfnamefont
  {S.}~\bibnamefont {Demura}}, \bibinfo {author} {\bibfnamefont
  {K.}~\bibnamefont {Deguchi}}, \ and\ \bibinfo {author} {\bibfnamefont
  {Y.}~\bibnamefont {Takano}},\ }\href {\doibase 10.1103/PhysRevB.89.214513}
  {\bibfield  {journal} {\bibinfo  {journal} {Phys. Rev. B}\ }\textbf {\bibinfo
  {volume} {89}},\ \bibinfo {pages} {1} (\bibinfo {year} {2014})}\BibitemShut
  {NoStop}%
\bibitem [{\citenamefont {Wolowiec}\ \emph {et~al.}(2013)\citenamefont
  {Wolowiec}, \citenamefont {Yazici}, \citenamefont {White}, \citenamefont
  {Huang},\ and\ \citenamefont {Maple}}]{Wolowiec2013}%
  \BibitemOpen
  \bibfield  {author} {\bibinfo {author} {\bibfnamefont {C.~T.}\ \bibnamefont
  {Wolowiec}}, \bibinfo {author} {\bibfnamefont {D.}~\bibnamefont {Yazici}},
  \bibinfo {author} {\bibfnamefont {B.~D.}\ \bibnamefont {White}}, \bibinfo
  {author} {\bibfnamefont {K.}~\bibnamefont {Huang}}, \ and\ \bibinfo {author}
  {\bibfnamefont {M.~B.}\ \bibnamefont {Maple}},\ }\href@noop {} {\bibfield
  {journal} {\bibinfo  {journal} {Phys. Rev. B}\ }\textbf {\bibinfo {volume}
  {88}},\ \bibinfo {pages} {064503} (\bibinfo {year} {2013})},\
  \BibitemShut {NoStop}%
\bibitem [{\citenamefont {Tomita}\ \emph {et~al.}(2014)\citenamefont {Tomita},
  \citenamefont {Ebata}, \citenamefont {Soeda}, \citenamefont {Takahashi},
  \citenamefont {Fujihisa}, \citenamefont {Gotoh}, \citenamefont {Mizuguchi},
  \citenamefont {Izawa}, \citenamefont {Miura}, \citenamefont {Demura},
  \citenamefont {Deguchi},\ and\ \citenamefont {Takano}}]{Tomita2014}%
  \BibitemOpen
  \bibfield  {author} {\bibinfo {author} {\bibfnamefont {T.}~\bibnamefont
  {Tomita}}, \bibinfo {author} {\bibfnamefont {M.}~\bibnamefont {Ebata}},
  \bibinfo {author} {\bibfnamefont {H.}~\bibnamefont {Soeda}}, \bibinfo
  {author} {\bibfnamefont {H.}~\bibnamefont {Takahashi}}, \bibinfo {author}
  {\bibfnamefont {H.}~\bibnamefont {Fujihisa}}, \bibinfo {author}
  {\bibfnamefont {Y.}~\bibnamefont {Gotoh}}, \bibinfo {author} {\bibfnamefont
  {Y.}~\bibnamefont {Mizuguchi}}, \bibinfo {author} {\bibfnamefont
  {H.}~\bibnamefont {Izawa}}, \bibinfo {author} {\bibfnamefont
  {O.}~\bibnamefont {Miura}}, \bibinfo {author} {\bibfnamefont
  {S.}~\bibnamefont {Demura}}, \bibinfo {author} {\bibfnamefont
  {K.}~\bibnamefont {Deguchi}}, \ and\ \bibinfo {author} {\bibfnamefont
  {Y.}~\bibnamefont {Takano}},\ }\href {\doibase 10.7566/JPSJ.83.063704}
  {\bibfield  {journal} {\bibinfo  {journal} {J. Phys. Soc. Japan}\ }\textbf
  {\bibinfo {volume} {83}},\ \bibinfo {pages} {063704} (\bibinfo {year}
  {2014})},\
  \BibitemShut {NoStop}%
\bibitem [{\citenamefont {Kotegawa}\ \emph {et~al.}(2012)\citenamefont
  {Kotegawa}, \citenamefont {Tomita}, \citenamefont {Tou}, \citenamefont
  {Izawa}, \citenamefont {Mizuguchi}, \citenamefont {Miura}, \citenamefont
  {Demura}, \citenamefont {Deguchi},\ and\ \citenamefont
  {Takano}}]{Kotegawa2012}%
  \BibitemOpen
  \bibfield  {author} {\bibinfo {author} {\bibfnamefont {H.}~\bibnamefont
  {Kotegawa}}, \bibinfo {author} {\bibfnamefont {Y.}~\bibnamefont {Tomita}},
  \bibinfo {author} {\bibfnamefont {H.}~\bibnamefont {Tou}}, \bibinfo {author}
  {\bibfnamefont {H.}~\bibnamefont {Izawa}}, \bibinfo {author} {\bibfnamefont
  {Y.}~\bibnamefont {Mizuguchi}}, \bibinfo {author} {\bibfnamefont
  {O.}~\bibnamefont {Miura}}, \bibinfo {author} {\bibfnamefont
  {S.}~\bibnamefont {Demura}}, \bibinfo {author} {\bibfnamefont
  {K.}~\bibnamefont {Deguchi}}, \ and\ \bibinfo {author} {\bibfnamefont
  {Y.}~\bibnamefont {Takano}},\ }\href@noop {} {\bibfield  {journal} {\bibinfo
  {journal} {J. Phys. Soc. Japan}\ }\textbf {\bibinfo {volume} {81}},\ \bibinfo
  {pages} {103702} (\bibinfo {year} {2012})}\BibitemShut {NoStop}%
\bibitem [{\citenamefont {Lee}\ \emph {et~al.}(2013)\citenamefont {Lee},
  \citenamefont {Stone}, \citenamefont {Huq}, \citenamefont {Yildirim},
  \citenamefont {Ehlers}, \citenamefont {Mizuguchi}, \citenamefont {Miura},
  \citenamefont {Takano}, \citenamefont {Deguchi}, \citenamefont {Demura},\
  and\ \citenamefont {Lee}}]{Lee2013}%
  \BibitemOpen
  \bibfield  {author} {\bibinfo {author} {\bibfnamefont {J.}~\bibnamefont
  {Lee}}, \bibinfo {author} {\bibfnamefont {M.~B.}\ \bibnamefont {Stone}},
  \bibinfo {author} {\bibfnamefont {A.}~\bibnamefont {Huq}}, \bibinfo {author}
  {\bibfnamefont {T.}~\bibnamefont {Yildirim}}, \bibinfo {author}
  {\bibfnamefont {G.}~\bibnamefont {Ehlers}}, \bibinfo {author} {\bibfnamefont
  {Y.}~\bibnamefont {Mizuguchi}}, \bibinfo {author} {\bibfnamefont
  {O.}~\bibnamefont {Miura}}, \bibinfo {author} {\bibfnamefont
  {Y.}~\bibnamefont {Takano}}, \bibinfo {author} {\bibfnamefont
  {K.}~\bibnamefont {Deguchi}}, \bibinfo {author} {\bibfnamefont
  {S.}~\bibnamefont {Demura}}, \ and\ \bibinfo {author} {\bibfnamefont {S.~H.}\
  \bibnamefont {Lee}},\ }\href {\doibase 10.1103/PhysRevB.87.205134} {\bibfield
   {journal} {\bibinfo  {journal} {Phys. Rev. B}\ }\textbf {\bibinfo {volume}
  {87}},\ \bibinfo {pages} {205134} (\bibinfo {year} {2013})},\ 
  \BibitemShut {NoStop}%
\bibitem [{\citenamefont {Athauda}\ \emph {et~al.}(2015)\citenamefont
  {Athauda}, \citenamefont {Yang}, \citenamefont {Lee}, \citenamefont
  {Mizuguchi}, \citenamefont {Deguchi}, \citenamefont {Takano}, \citenamefont
  {Miura},\ and\ \citenamefont {Louca}}]{Athauda2015}%
  \BibitemOpen
  \bibfield  {author} {\bibinfo {author} {\bibfnamefont {A.}~\bibnamefont
  {Athauda}}, \bibinfo {author} {\bibfnamefont {J.}~\bibnamefont {Yang}},
  \bibinfo {author} {\bibfnamefont {S.}~\bibnamefont {Lee}}, \bibinfo {author}
  {\bibfnamefont {Y.}~\bibnamefont {Mizuguchi}}, \bibinfo {author}
  {\bibfnamefont {K.}~\bibnamefont {Deguchi}}, \bibinfo {author} {\bibfnamefont
  {Y.}~\bibnamefont {Takano}}, \bibinfo {author} {\bibfnamefont
  {O.}~\bibnamefont {Miura}}, \ and\ \bibinfo {author} {\bibfnamefont
  {D.}~\bibnamefont {Louca}},\ }\href {\doibase 10.1103/PhysRevB.91.144112}
  {\bibfield  {journal} {\bibinfo  {journal} {Phys. Rev. B}\ }\textbf {\bibinfo
  {volume} {91}},\ \bibinfo {pages} {1} (\bibinfo {year} {2015})},\ 
  \BibitemShut {NoStop}%
\bibitem [{\citenamefont {Zeng}\ \emph {et~al.}(2014)\citenamefont {Zeng},
  \citenamefont {Wang}, \citenamefont {Ma}, \citenamefont {Richard},
  \citenamefont {Nie}, \citenamefont {Weng}, \citenamefont {Wang},
  \citenamefont {Wang}, \citenamefont {Qian},\ and\ \citenamefont
  {Ding}}]{Zeng2014}%
  \BibitemOpen
  \bibfield  {author} {\bibinfo {author} {\bibfnamefont {L.~K.}\ \bibnamefont
  {Zeng}}, \bibinfo {author} {\bibfnamefont {X.~B.}\ \bibnamefont {Wang}},
  \bibinfo {author} {\bibfnamefont {J.}~\bibnamefont {Ma}}, \bibinfo {author}
  {\bibfnamefont {P.}~\bibnamefont {Richard}}, \bibinfo {author} {\bibfnamefont
  {S.~M.}\ \bibnamefont {Nie}}, \bibinfo {author} {\bibfnamefont {H.~M.}\
  \bibnamefont {Weng}}, \bibinfo {author} {\bibfnamefont {N.~L.}\ \bibnamefont
  {Wang}}, \bibinfo {author} {\bibfnamefont {Z.}~\bibnamefont {Wang}}, \bibinfo
  {author} {\bibfnamefont {T.}~\bibnamefont {Qian}}, \ and\ \bibinfo {author}
  {\bibfnamefont {H.}~\bibnamefont {Ding}},\ }\href {\doibase
  10.1103/PhysRevB.90.054512} {\bibfield  {journal} {\bibinfo  {journal} {Phys.
  Rev. B}\ }\textbf {\bibinfo {volume} {90}},\ \bibinfo {pages} {1} (\bibinfo
  {year} {2014})},\  \BibitemShut {NoStop}%
\bibitem [{\citenamefont {Ye}\ \emph {et~al.}(2014)\citenamefont {Ye},
  \citenamefont {Yang}, \citenamefont {Shen}, \citenamefont {Jiang},
  \citenamefont {Niu}, \citenamefont {Feng}, \citenamefont {Du}, \citenamefont
  {Wan}, \citenamefont {Liu}, \citenamefont {Zhu}, \citenamefont {Wen},\ and\
  \citenamefont {Jiang}}]{Ye2014}%
  \BibitemOpen
  \bibfield  {author} {\bibinfo {author} {\bibfnamefont {Z.~R.}\ \bibnamefont
  {Ye}}, \bibinfo {author} {\bibfnamefont {H.~F.}\ \bibnamefont {Yang}},
  \bibinfo {author} {\bibfnamefont {D.~W.}\ \bibnamefont {Shen}}, \bibinfo
  {author} {\bibfnamefont {J.}~\bibnamefont {Jiang}}, \bibinfo {author}
  {\bibfnamefont {X.~H.}\ \bibnamefont {Niu}}, \bibinfo {author} {\bibfnamefont
  {D.~L.}\ \bibnamefont {Feng}}, \bibinfo {author} {\bibfnamefont {Y.~P.}\
  \bibnamefont {Du}}, \bibinfo {author} {\bibfnamefont {X.~G.}\ \bibnamefont
  {Wan}}, \bibinfo {author} {\bibfnamefont {J.~Z.}\ \bibnamefont {Liu}},
  \bibinfo {author} {\bibfnamefont {X.~Y.}\ \bibnamefont {Zhu}}, \bibinfo
  {author} {\bibfnamefont {H.-H.}\ \bibnamefont {Wen}}, \ and\ \bibinfo
  {author} {\bibfnamefont {M.~H.}\ \bibnamefont {Jiang}},\ }\href {\doibase
  10.1103/PhysRevB.90.045116} {\bibfield  {journal} {\bibinfo  {journal} {Phys.
  Rev. B}\ }\textbf {\bibinfo {volume} {90}},\ \bibinfo {pages} {045116}
  (\bibinfo {year} {2014})}\BibitemShut {NoStop}%
\bibitem [{\citenamefont {Terashima}\ \emph {et~al.}(2014)\citenamefont
  {Terashima}, \citenamefont {Sonoyama}, \citenamefont {Wakita}, \citenamefont
  {Sunagawa}, \citenamefont {Ono}, \citenamefont {Kumigashira}, \citenamefont
  {Muro}, \citenamefont {Nagao}, \citenamefont {Watauchi}, \citenamefont
  {Tanaka}, \citenamefont {Okazaki}, \citenamefont {Takano}, \citenamefont
  {Miura}, \citenamefont {Mizuguchi}, \citenamefont {Usui}, \citenamefont
  {Suzuki}, \citenamefont {Kuroki}, \citenamefont {Muraoka},\ and\
  \citenamefont {Yokoya}}]{Terashima2014}%
  \BibitemOpen
  \bibfield  {author} {\bibinfo {author} {\bibfnamefont {K.}~\bibnamefont
  {Terashima}}, \bibinfo {author} {\bibfnamefont {J.}~\bibnamefont {Sonoyama}},
  \bibinfo {author} {\bibfnamefont {T.}~\bibnamefont {Wakita}}, \bibinfo
  {author} {\bibfnamefont {M.}~\bibnamefont {Sunagawa}}, \bibinfo {author}
  {\bibfnamefont {K.}~\bibnamefont {Ono}}, \bibinfo {author} {\bibfnamefont
  {H.}~\bibnamefont {Kumigashira}}, \bibinfo {author} {\bibfnamefont
  {T.}~\bibnamefont {Muro}}, \bibinfo {author} {\bibfnamefont {M.}~\bibnamefont
  {Nagao}}, \bibinfo {author} {\bibfnamefont {S.}~\bibnamefont {Watauchi}},
  \bibinfo {author} {\bibfnamefont {I.}~\bibnamefont {Tanaka}}, \bibinfo
  {author} {\bibfnamefont {H.}~\bibnamefont {Okazaki}}, \bibinfo {author}
  {\bibfnamefont {Y.}~\bibnamefont {Takano}}, \bibinfo {author} {\bibfnamefont
  {O.}~\bibnamefont {Miura}}, \bibinfo {author} {\bibfnamefont
  {Y.}~\bibnamefont {Mizuguchi}}, \bibinfo {author} {\bibfnamefont
  {H.}~\bibnamefont {Usui}}, \bibinfo {author} {\bibfnamefont {K.}~\bibnamefont
  {Suzuki}}, \bibinfo {author} {\bibfnamefont {K.}~\bibnamefont {Kuroki}},
  \bibinfo {author} {\bibfnamefont {Y.}~\bibnamefont {Muraoka}}, \ and\
  \bibinfo {author} {\bibfnamefont {T.}~\bibnamefont {Yokoya}},\ }\href
  {\doibase 10.1103/PhysRevB.90.220512} {\bibfield  {journal} {\bibinfo
  {journal} {Phys. Rev. B}\ }\textbf {\bibinfo {volume} {90}},\ \bibinfo
  {pages} {1} (\bibinfo {year} {2014})},\  \BibitemShut {NoStop}%
\bibitem [{\citenamefont {Sugimoto}\ \emph {et~al.}(2015)\citenamefont
  {Sugimoto}, \citenamefont {Ootsuki}, \citenamefont {Morice}, \citenamefont
  {Artacho}, \citenamefont {Saxena}, \citenamefont {Schwier}, \citenamefont
  {Zheng}, \citenamefont {Kojima}, \citenamefont {Iwasawa}, \citenamefont
  {Shimada}, \citenamefont {Arita}, \citenamefont {Namatame}, \citenamefont
  {Taniguchi}, \citenamefont {Takahashi}, \citenamefont {Saini}, \citenamefont
  {Asano}, \citenamefont {Higashinaka}, \citenamefont {Matsuda}, \citenamefont
  {Aoki},\ and\ \citenamefont {Mizokawa}}]{Sugimoto2015}%
  \BibitemOpen
  \bibfield  {author} {\bibinfo {author} {\bibfnamefont {T.}~\bibnamefont
  {Sugimoto}}, \bibinfo {author} {\bibfnamefont {D.}~\bibnamefont {Ootsuki}},
  \bibinfo {author} {\bibfnamefont {C.}~\bibnamefont {Morice}}, \bibinfo
  {author} {\bibfnamefont {E.}~\bibnamefont {Artacho}}, \bibinfo {author}
  {\bibfnamefont {S.~S.}\ \bibnamefont {Saxena}}, \bibinfo {author}
  {\bibfnamefont {E.~F.}\ \bibnamefont {Schwier}}, \bibinfo {author}
  {\bibfnamefont {M.}~\bibnamefont {Zheng}}, \bibinfo {author} {\bibfnamefont
  {Y.}~\bibnamefont {Kojima}}, \bibinfo {author} {\bibfnamefont
  {H.}~\bibnamefont {Iwasawa}}, \bibinfo {author} {\bibfnamefont
  {K.}~\bibnamefont {Shimada}}, \bibinfo {author} {\bibfnamefont
  {M.}~\bibnamefont {Arita}}, \bibinfo {author} {\bibfnamefont
  {H.}~\bibnamefont {Namatame}}, \bibinfo {author} {\bibfnamefont
  {M.}~\bibnamefont {Taniguchi}}, \bibinfo {author} {\bibfnamefont
  {M.}~\bibnamefont {Takahashi}}, \bibinfo {author} {\bibfnamefont {N.~L.}\
  \bibnamefont {Saini}}, \bibinfo {author} {\bibfnamefont {T.}~\bibnamefont
  {Asano}}, \bibinfo {author} {\bibfnamefont {R.}~\bibnamefont {Higashinaka}},
  \bibinfo {author} {\bibfnamefont {T.~D.}\ \bibnamefont {Matsuda}}, \bibinfo
  {author} {\bibfnamefont {Y.}~\bibnamefont {Aoki}}, \ and\ \bibinfo {author}
  {\bibfnamefont {T.}~\bibnamefont {Mizokawa}},\ }\href {\doibase
  10.1103/PhysRevB.92.041113} {\bibfield  {journal} {\bibinfo  {journal} {Phys.
  Rev. B}\ }\textbf {\bibinfo {volume} {92}},\ \bibinfo {pages} {041113}
  (\bibinfo {year} {2015})},\ \BibitemShut {NoStop}%
\bibitem [{\citenamefont {Martins}\ \emph {et~al.}(2013)\citenamefont
  {Martins}, \citenamefont {Moreo},\ and\ \citenamefont
  {Dagotto}}]{Martins2013}%
  \BibitemOpen
  \bibfield  {author} {\bibinfo {author} {\bibfnamefont {G.~B.}\ \bibnamefont
  {Martins}}, \bibinfo {author} {\bibfnamefont {A.}~\bibnamefont {Moreo}}, \
  and\ \bibinfo {author} {\bibfnamefont {E.}~\bibnamefont {Dagotto}},\ }\href
  {\doibase 10.1103/PhysRevB.87.081102} {\bibfield  {journal} {\bibinfo
  {journal} {Phys. Rev. B}\ }\textbf {\bibinfo {volume} {87}},\ \bibinfo
  {pages} {81102} (\bibinfo {year} {2013})}\BibitemShut {NoStop}%
\bibitem [{\citenamefont {Zhou}\ and\ \citenamefont {Wang}(2013)}]{Zhou2013}%
  \BibitemOpen
  \bibfield  {author} {\bibinfo {author} {\bibfnamefont {T.}~\bibnamefont
  {Zhou}}\ and\ \bibinfo {author} {\bibfnamefont {Z.~D.}\ \bibnamefont
  {Wang}},\ }\href {\doibase 10.1007/s10948-012-2073-4} {\bibfield  {journal}
  {\bibinfo  {journal} {J. Supercond. Nov. Magn.}\ }\textbf {\bibinfo {volume}
  {26}},\ \bibinfo {pages} {2735} (\bibinfo {year} {2013})}\BibitemShut
  {NoStop}%
\bibitem [{\citenamefont {Wu}\ \emph {et~al.}(2014)\citenamefont {Wu},
  \citenamefont {Yuan}, \citenamefont {Liang}, \citenamefont {Fan},\ and\
  \citenamefont {Hu}}]{Wu2014}%
  \BibitemOpen
  \bibfield  {author} {\bibinfo {author} {\bibfnamefont {X.}~\bibnamefont
  {Wu}}, \bibinfo {author} {\bibfnamefont {J.}~\bibnamefont {Yuan}}, \bibinfo
  {author} {\bibfnamefont {Y.}~\bibnamefont {Liang}}, \bibinfo {author}
  {\bibfnamefont {H.}~\bibnamefont {Fan}}, \ and\ \bibinfo {author}
  {\bibfnamefont {J.}~\bibnamefont {Hu}},\ }\href
  {http://stacks.iop.org/0295-5075/108/i=2/a=27006} {\bibfield  {journal}
  {\bibinfo  {journal} {Europhys. Lett.}\ }\textbf {\bibinfo {volume} {108}},\
  \bibinfo {pages} {27006} (\bibinfo {year} {2014})}\BibitemShut {NoStop}%
\bibitem [{\citenamefont {Liang}\ \emph {et~al.}(2014)\citenamefont {Liang},
  \citenamefont {Wu}, \citenamefont {Tsai},\ and\ \citenamefont
  {Hu}}]{Liang2014}%
  \BibitemOpen
  \bibfield  {author} {\bibinfo {author} {\bibfnamefont {Y.}~\bibnamefont
  {Liang}}, \bibinfo {author} {\bibfnamefont {X.}~\bibnamefont {Wu}}, \bibinfo
  {author} {\bibfnamefont {W.-F.}\ \bibnamefont {Tsai}}, \ and\ \bibinfo
  {author} {\bibfnamefont {J.}~\bibnamefont {Hu}},\ }\href {\doibase
  10.1007/s11467-013-0407-8} {\bibfield  {journal} {\bibinfo  {journal} {Front.
  Phys.}\ }\textbf {\bibinfo {volume} {9}},\ \bibinfo {pages} {194} (\bibinfo
  {year} {2014})}\BibitemShut {NoStop}%
\bibitem [{\citenamefont {Liu}\ and\ \citenamefont {Feng}(2014)}]{Liu2014}%
  \BibitemOpen
  \bibfield  {author} {\bibinfo {author} {\bibfnamefont {B.}~\bibnamefont
  {Liu}}\ and\ \bibinfo {author} {\bibfnamefont {S.}~\bibnamefont {Feng}},\
  }\href {http://stacks.iop.org/0295-5075/106/i=1/a=17003} {\bibfield
  {journal} {\bibinfo  {journal} {Europhys. Lett.}\ }\textbf {\bibinfo {volume}
  {106}},\ \bibinfo {pages} {17003} (\bibinfo {year} {2014})}\BibitemShut
  {NoStop}%
\bibitem [{\citenamefont {Yang}\ \emph {et~al.}(2013)\citenamefont {Yang},
  \citenamefont {Wang}, \citenamefont {Xiang}, \citenamefont {Li},\ and\
  \citenamefont {Wang}}]{Yang2013}%
  \BibitemOpen
  \bibfield  {author} {\bibinfo {author} {\bibfnamefont {Y.}~\bibnamefont
  {Yang}}, \bibinfo {author} {\bibfnamefont {W.~S.}\ \bibnamefont {Wang}},
  \bibinfo {author} {\bibfnamefont {Y.~Y.}\ \bibnamefont {Xiang}}, \bibinfo
  {author} {\bibfnamefont {Z.~Z.}\ \bibnamefont {Li}}, \ and\ \bibinfo {author}
  {\bibfnamefont {Q.-H.}\ \bibnamefont {Wang}},\ }\href {\doibase
  10.1103/PhysRevB.88.094519} {\bibfield  {journal} {\bibinfo  {journal} {Phys.
  Rev. B}\ }\textbf {\bibinfo {volume} {88}},\ \bibinfo {pages} {1} (\bibinfo
  {year} {2013})},\ 
  \BibitemShut {NoStop}%
\bibitem [{\citenamefont {Li}\ \emph {et~al.}(2013)\citenamefont {Li},
  \citenamefont {Xing},\ and\ \citenamefont {Huang}}]{Li2013}%
  \BibitemOpen
  \bibfield  {author} {\bibinfo {author} {\bibfnamefont {B.}~\bibnamefont
  {Li}}, \bibinfo {author} {\bibfnamefont {Z.~W.}\ \bibnamefont {Xing}}, \ and\
  \bibinfo {author} {\bibfnamefont {G.~Q.}\ \bibnamefont {Huang}},\ }\href
  {http://stacks.iop.org/0295-5075/101/i=4/a=47002} {\bibfield  {journal}
  {\bibinfo  {journal} {Europhys. Lett.}\ }\textbf {\bibinfo {volume} {101}},\
  \bibinfo {pages} {47002} (\bibinfo {year} {2013})}\BibitemShut {NoStop}%
\bibitem [{\citenamefont {Feng}\ \emph {et~al.}(2014)\citenamefont {Feng},
  \citenamefont {Ding}, \citenamefont {Du}, \citenamefont {Wan}, \citenamefont
  {Wang}, \citenamefont {Savrasov},\ and\ \citenamefont {Duan}}]{Feng2014a}%
  \BibitemOpen
  \bibfield  {author} {\bibinfo {author} {\bibfnamefont {Y.}~\bibnamefont
  {Feng}}, \bibinfo {author} {\bibfnamefont {H.-C.}\ \bibnamefont {Ding}},
  \bibinfo {author} {\bibfnamefont {Y.}~\bibnamefont {Du}}, \bibinfo {author}
  {\bibfnamefont {X.}~\bibnamefont {Wan}}, \bibinfo {author} {\bibfnamefont
  {B.}~\bibnamefont {Wang}}, \bibinfo {author} {\bibfnamefont {S.~Y.}\
  \bibnamefont {Savrasov}}, \ and\ \bibinfo {author} {\bibfnamefont {C.-G.}\
  \bibnamefont {Duan}},\ }\href {\doibase 10.1063/1.4883755} {\bibfield
  {journal} {\bibinfo  {journal} {J. Appl. Phys.}\ }\textbf {\bibinfo {volume}
  {115}},\ \bibinfo {pages} {5} (\bibinfo {year} {2014})},\ \BibitemShut {NoStop}%
\bibitem [{\citenamefont {L{\"{u}}ders}\ \emph {et~al.}(2005)\citenamefont
  {L{\"{u}}ders}, \citenamefont {Marques}, \citenamefont {Lathiotakis},
  \citenamefont {Floris}, \citenamefont {Profeta}, \citenamefont {Fast},
  \citenamefont {Continenza}, \citenamefont {Massidda},\ and\ \citenamefont
  {Gross}}]{Luders2005}%
  \BibitemOpen
  \bibfield  {author} {\bibinfo {author} {\bibfnamefont {M.}~\bibnamefont
  {L{\"{u}}ders}}, \bibinfo {author} {\bibfnamefont {M.~A.~L.}\ \bibnamefont
  {Marques}}, \bibinfo {author} {\bibfnamefont {N.~N.}\ \bibnamefont
  {Lathiotakis}}, \bibinfo {author} {\bibfnamefont {A.}~\bibnamefont {Floris}},
  \bibinfo {author} {\bibfnamefont {G.}~\bibnamefont {Profeta}}, \bibinfo
  {author} {\bibfnamefont {L.}~\bibnamefont {Fast}}, \bibinfo {author}
  {\bibfnamefont {A.}~\bibnamefont {Continenza}}, \bibinfo {author}
  {\bibfnamefont {S.}~\bibnamefont {Massidda}}, \ and\ \bibinfo {author}
  {\bibfnamefont {E.~K.~U.}\ \bibnamefont {Gross}},\ }\href {\doibase
  10.1103/PhysRevB.72.024545} {\bibfield  {journal} {\bibinfo  {journal} {Phys.
  Rev. B}\ }\textbf {\bibinfo {volume} {72}},\ \bibinfo {pages} {1} (\bibinfo
  {year} {2005})},\ \BibitemShut {NoStop}%
\bibitem [{\citenamefont {Marques}\ \emph {et~al.}(2005)\citenamefont
  {Marques}, \citenamefont {L{\"{u}}ders}, \citenamefont {Lathiotakis},
  \citenamefont {Profeta}, \citenamefont {Floris}, \citenamefont {Fast},
  \citenamefont {Continenza}, \citenamefont {Gross},\ and\ \citenamefont
  {Massidda}}]{Marques2005a}%
  \BibitemOpen
  \bibfield  {author} {\bibinfo {author} {\bibfnamefont {M.~A.~L.}\
  \bibnamefont {Marques}}, \bibinfo {author} {\bibfnamefont {M.}~\bibnamefont
  {L{\"{u}}ders}}, \bibinfo {author} {\bibfnamefont {N.~N.}\ \bibnamefont
  {Lathiotakis}}, \bibinfo {author} {\bibfnamefont {G.}~\bibnamefont
  {Profeta}}, \bibinfo {author} {\bibfnamefont {A.}~\bibnamefont {Floris}},
  \bibinfo {author} {\bibfnamefont {L.}~\bibnamefont {Fast}}, \bibinfo {author}
  {\bibfnamefont {A.}~\bibnamefont {Continenza}}, \bibinfo {author}
  {\bibfnamefont {E.~K.~U.}\ \bibnamefont {Gross}}, \ and\ \bibinfo {author}
  {\bibfnamefont {S.}~\bibnamefont {Massidda}},\ }\href {\doibase
  10.1103/PhysRevB.72.024546} {\bibfield  {journal} {\bibinfo  {journal} {Phys.
  Rev. B}\ }\textbf {\bibinfo {volume} {72}},\ \bibinfo {pages} {1} (\bibinfo
  {year} {2005})},\ \BibitemShut {NoStop}%
\bibitem [{\citenamefont {Giannozzi}\ \emph {et~al.}(2009)\citenamefont
  {Giannozzi}, \citenamefont {Baroni}, \citenamefont {Bonini}, \citenamefont
  {Calandra}, \citenamefont {Car}, \citenamefont {Cavazzoni}, \citenamefont
  {Ceresoli}, \citenamefont {Chiarotti}, \citenamefont {Cococcioni},
  \citenamefont {Dabo}, \citenamefont {{Dal Corso}}, \citenamefont
  {de~Gironcoli}, \citenamefont {Fabris}, \citenamefont {Fratesi},
  \citenamefont {Gebauer}, \citenamefont {Gerstmann}, \citenamefont
  {Gougoussis}, \citenamefont {Kokalj}, \citenamefont {Lazzeri}, \citenamefont
  {Martin-Samos}, \citenamefont {Marzari}, \citenamefont {Mauri}, \citenamefont
  {Mazzarello}, \citenamefont {Paolini}, \citenamefont {Pasquarello},
  \citenamefont {Paulatto}, \citenamefont {Sbraccia}, \citenamefont {Scandolo},
  \citenamefont {Sclauzero}, \citenamefont {Seitsonen}, \citenamefont
  {Smogunov}, \citenamefont {Umari},\ and\ \citenamefont
  {Wentzcovitch}}]{Giannozzi2009}%
  \BibitemOpen
  \bibfield  {author} {\bibinfo {author} {\bibfnamefont {P.}~\bibnamefont
  {Giannozzi}}, \bibinfo {author} {\bibfnamefont {S.}~\bibnamefont {Baroni}},
  \bibinfo {author} {\bibfnamefont {N.}~\bibnamefont {Bonini}}, \bibinfo
  {author} {\bibfnamefont {M.}~\bibnamefont {Calandra}}, \bibinfo {author}
  {\bibfnamefont {R.}~\bibnamefont {Car}}, \bibinfo {author} {\bibfnamefont
  {C.}~\bibnamefont {Cavazzoni}}, \bibinfo {author} {\bibfnamefont
  {D.}~\bibnamefont {Ceresoli}}, \bibinfo {author} {\bibfnamefont {G.~L.}\
  \bibnamefont {Chiarotti}}, \bibinfo {author} {\bibfnamefont {M.}~\bibnamefont
  {Cococcioni}}, \bibinfo {author} {\bibfnamefont {I.}~\bibnamefont {Dabo}},
  \bibinfo {author} {\bibfnamefont {A.}~\bibnamefont {{Dal Corso}}}, \bibinfo
  {author} {\bibfnamefont {S.}~\bibnamefont {de~Gironcoli}}, \bibinfo {author}
  {\bibfnamefont {S.}~\bibnamefont {Fabris}}, \bibinfo {author} {\bibfnamefont
  {G.}~\bibnamefont {Fratesi}}, \bibinfo {author} {\bibfnamefont
  {R.}~\bibnamefont {Gebauer}}, \bibinfo {author} {\bibfnamefont
  {U.}~\bibnamefont {Gerstmann}}, \bibinfo {author} {\bibfnamefont
  {C.}~\bibnamefont {Gougoussis}}, \bibinfo {author} {\bibfnamefont
  {A.}~\bibnamefont {Kokalj}}, \bibinfo {author} {\bibfnamefont
  {M.}~\bibnamefont {Lazzeri}}, \bibinfo {author} {\bibfnamefont
  {L.}~\bibnamefont {Martin-Samos}}, \bibinfo {author} {\bibfnamefont
  {N.}~\bibnamefont {Marzari}}, \bibinfo {author} {\bibfnamefont
  {F.}~\bibnamefont {Mauri}}, \bibinfo {author} {\bibfnamefont
  {R.}~\bibnamefont {Mazzarello}}, \bibinfo {author} {\bibfnamefont
  {S.}~\bibnamefont {Paolini}}, \bibinfo {author} {\bibfnamefont
  {A.}~\bibnamefont {Pasquarello}}, \bibinfo {author} {\bibfnamefont
  {L.}~\bibnamefont {Paulatto}}, \bibinfo {author} {\bibfnamefont
  {C.}~\bibnamefont {Sbraccia}}, \bibinfo {author} {\bibfnamefont
  {S.}~\bibnamefont {Scandolo}}, \bibinfo {author} {\bibfnamefont
  {G.}~\bibnamefont {Sclauzero}}, \bibinfo {author} {\bibfnamefont {A.~P.}\
  \bibnamefont {Seitsonen}}, \bibinfo {author} {\bibfnamefont {A.}~\bibnamefont
  {Smogunov}}, \bibinfo {author} {\bibfnamefont {P.}~\bibnamefont {Umari}}, \
  and\ \bibinfo {author} {\bibfnamefont {R.~M.}\ \bibnamefont {Wentzcovitch}},\
  }\href {\doibase 10.1088/0953-8984/21/39/395502} {\bibfield  {journal}
  {\bibinfo  {journal} {J. Phys. Condens. Matter}\ }\textbf {\bibinfo {volume}
  {21}},\ \bibinfo {pages} {395502} (\bibinfo {year} {2009})},\ \BibitemShut {NoStop}%
\bibitem [{Elk(2004)}]{Elk}%
  \BibitemOpen
  \href@noop {} {\enquote {\bibinfo {title} {http://elk.sourceforge.net},}\ }
  (\bibinfo {year} {2004})\BibitemShut {NoStop}%
\bibitem [{\citenamefont {Perdew}\ and\ \citenamefont
  {Zunger}(1981)}]{Perdew1981}%
  \BibitemOpen
  \bibfield  {author} {\bibinfo {author} {\bibfnamefont {J.~P.}\ \bibnamefont
  {Perdew}}\ and\ \bibinfo {author} {\bibfnamefont {A.}~\bibnamefont
  {Zunger}},\ }\href {\doibase 10.1103/PhysRevB.23.5048} {\bibfield  {journal}
  {\bibinfo  {journal} {Phys. Rev. B}\ }\textbf {\bibinfo {volume} {23}},\
  \bibinfo {pages} {5048} (\bibinfo {year} {1981})},\ \BibitemShut {NoStop}%
\bibitem [{\citenamefont {Koelling}\ and\ \citenamefont
  {Harmon}(1977)}]{Koelling1977a}%
  \BibitemOpen
  \bibfield  {author} {\bibinfo {author} {\bibfnamefont {D.~D.}\ \bibnamefont
  {Koelling}}\ and\ \bibinfo {author} {\bibfnamefont {B.~N.}\ \bibnamefont
  {Harmon}},\ }\href {\doibase 10.1088/0022-3719/10/16/019} {\bibfield
  {journal} {\bibinfo  {journal} {J. Phys. C Solid State Phys.}\ }\textbf
  {\bibinfo {volume} {10}},\ \bibinfo {pages} {3107} (\bibinfo {year}
  {1977})}\BibitemShut {NoStop}%
\bibitem [{\citenamefont {Migdal}(1958)}]{Migdal1958}%
  \BibitemOpen
  \bibfield  {author} {\bibinfo {author} {\bibfnamefont {A.}~\bibnamefont
  {Migdal}},\ }\href@noop {} {\bibfield  {journal} {\bibinfo  {journal} {JETP
  Lett.}\ }\textbf {\bibinfo {volume} {34}},\ \bibinfo {pages} {996} (\bibinfo
  {year} {1958})}\BibitemShut {NoStop}%
\bibitem [{\citenamefont {Eliashberg}(1960)}]{Eliashberg1960}%
  \BibitemOpen
  \bibfield  {author} {\bibinfo {author} {\bibfnamefont {G.~M.}\ \bibnamefont
  {Eliashberg}},\ }\href@noop {} {\bibfield  {journal} {\bibinfo  {journal}
  {Sov. Phys. JETP}\ }\textbf {\bibinfo {volume} {11}},\ \bibinfo {pages} {696}
  (\bibinfo {year} {1960})}\BibitemShut {NoStop}%
\bibitem [{\citenamefont {Eliashberg}(1961)}]{Eliashberg1961}%
  \BibitemOpen
  \bibfield  {author} {\bibinfo {author} {\bibfnamefont {G.~M.}\ \bibnamefont
  {Eliashberg}},\ }\href@noop {} {\bibfield  {journal} {\bibinfo  {journal}
  {Sov. Phys. JETP}\ }\textbf {\bibinfo {volume} {12}},\ \bibinfo {pages}
  {1000} (\bibinfo {year} {1961})}\BibitemShut {NoStop}%
\bibitem [{\citenamefont {Mermin}(1965)}]{Mermin1965}%
  \BibitemOpen
  \bibfield  {author} {\bibinfo {author} {\bibfnamefont {N.~D.}\ \bibnamefont
  {Mermin}},\ }\href {\doibase 10.1103/PhysRev.137.A1441} {\bibfield  {journal}
  {\bibinfo  {journal} {Phys. Rev.}\ }\textbf {\bibinfo {volume} {137}},\
  \bibinfo {pages} {A1441} (\bibinfo {year} {1965})}\BibitemShut {NoStop}%
\bibitem [{\citenamefont {G{\"{o}}rling}\ and\ \citenamefont
  {Levy}(1994)}]{Gorling1994}%
  \BibitemOpen
  \bibfield  {author} {\bibinfo {author} {\bibfnamefont {A.}~\bibnamefont
  {G{\"{o}}rling}}\ and\ \bibinfo {author} {\bibfnamefont {M.}~\bibnamefont
  {Levy}},\ }\href {\doibase 10.1103/PhysRevA.50.196} {\bibfield  {journal}
  {\bibinfo  {journal} {Phys. Rev. A}\ }\textbf {\bibinfo {volume} {50}},\
  \bibinfo {pages} {196} (\bibinfo {year} {1994})}\BibitemShut {NoStop}%
\bibitem [{\citenamefont {Akashi}\ and\ \citenamefont
  {Arita}(2013{\natexlab{a}})}]{Akashi2013a}%
  \BibitemOpen
  \bibfield  {author} {\bibinfo {author} {\bibfnamefont {R.}~\bibnamefont
  {Akashi}}\ and\ \bibinfo {author} {\bibfnamefont {R.}~\bibnamefont {Arita}},\
  }\href {\doibase 10.1103/PhysRevB.88.014514} {\bibfield  {journal} {\bibinfo
  {journal} {Phys. Rev. B}\ }\textbf {\bibinfo {volume} {88}},\ \bibinfo
  {pages} {14514} (\bibinfo {year} {2013}{\natexlab{a}})}\BibitemShut {NoStop}%
\bibitem [{\citenamefont {Baroni}\ \emph {et~al.}(2001)\citenamefont {Baroni},
  \citenamefont {de~Gironcoli}, \citenamefont {{Dal Corso}},\ and\
  \citenamefont {Giannozzi}}]{Baroni2001}%
  \BibitemOpen
  \bibfield  {author} {\bibinfo {author} {\bibfnamefont {S.}~\bibnamefont
  {Baroni}}, \bibinfo {author} {\bibfnamefont {S.}~\bibnamefont
  {de~Gironcoli}}, \bibinfo {author} {\bibfnamefont {A.}~\bibnamefont {{Dal
  Corso}}}, \ and\ \bibinfo {author} {\bibfnamefont {P.}~\bibnamefont
  {Giannozzi}},\ }\href {\doibase 10.1103/RevModPhys.73.515} {\bibfield
  {journal} {\bibinfo  {journal} {Rev. Mod. Phys.}\ }\textbf {\bibinfo {volume}
  {73}},\ \bibinfo {pages} {515} (\bibinfo {year} {2001})}\BibitemShut
  {NoStop}%
\bibitem [{\citenamefont {Pickett}(1982)}]{Pickett1982}%
  \BibitemOpen
  \bibfield  {author} {\bibinfo {author} {\bibfnamefont {W.~E.}\ \bibnamefont
  {Pickett}},\ }\href {\doibase 10.1103/PhysRevB.26.1186} {\bibfield  {journal}
  {\bibinfo  {journal} {Phys. Rev. B}\ }\textbf {\bibinfo {volume} {26}},\
  \bibinfo {pages} {1186} (\bibinfo {year} {1982})}\BibitemShut {NoStop}%
\bibitem [{\citenamefont {Akashi}\ and\ \citenamefont
  {Arita}(2013{\natexlab{b}})}]{Akashi2013}%
  \BibitemOpen
  \bibfield  {author} {\bibinfo {author} {\bibfnamefont {R.}~\bibnamefont
  {Akashi}}\ and\ \bibinfo {author} {\bibfnamefont {R.}~\bibnamefont {Arita}},\
  }\href {\doibase 10.1103/PhysRevLett.111.057006} {\bibfield  {journal}
  {\bibinfo  {journal} {Phys. Rev. Lett.}\ }\textbf {\bibinfo {volume} {111}},\
  \bibinfo {pages} {57006} (\bibinfo {year} {2013}{\natexlab{b}})}\BibitemShut
  {NoStop}%
\bibitem [{\citenamefont {Akashi}\ and\ \citenamefont
  {Arita}(2014)}]{Akashi2013c}%
  \BibitemOpen
  \bibfield  {author} {\bibinfo {author} {\bibfnamefont {R.}~\bibnamefont
  {Akashi}}\ and\ \bibinfo {author} {\bibfnamefont {R.}~\bibnamefont {Arita}},\
  }\href@noop {} {\bibfield  {journal} {\bibinfo  {journal} {J. Phys. Soc.
  Japan}\ }\textbf {\bibinfo {volume} {83}},\ \bibinfo {pages} {061016}
  (\bibinfo {year} {2014})},\ \BibitemShut {NoStop}%
\bibitem [{\citenamefont {Akashi}\ \emph {et~al.}(2015)\citenamefont {Akashi},
  \citenamefont {Kawamura}, \citenamefont {Tsuneyuki}, \citenamefont {Nomura},\
  and\ \citenamefont {Arita}}]{Akashi2015}%
  \BibitemOpen
  \bibfield  {author} {\bibinfo {author} {\bibfnamefont {R.}~\bibnamefont
  {Akashi}}, \bibinfo {author} {\bibfnamefont {M.}~\bibnamefont {Kawamura}},
  \bibinfo {author} {\bibfnamefont {S.}~\bibnamefont {Tsuneyuki}}, \bibinfo
  {author} {\bibfnamefont {Y.}~\bibnamefont {Nomura}}, \ and\ \bibinfo {author}
  {\bibfnamefont {R.}~\bibnamefont {Arita}},\ }\href {\doibase
  10.1103/PhysRevB.91.224513} {\bibfield  {journal} {\bibinfo  {journal} {Phys.
  Rev. B}\ }\textbf {\bibinfo {volume} {91}},\ \bibinfo {pages} {224513}
  (\bibinfo {year} {2015})},\ \BibitemShut {NoStop}%
\bibitem [{\citenamefont {Takada}(1978)}]{Takada1978}%
  \BibitemOpen
  \bibfield  {author} {\bibinfo {author} {\bibfnamefont {Y.}~\bibnamefont
  {Takada}},\ }\href {\doibase 10.1143/JPSJ.45.786} {\bibfield  {journal}
  {\bibinfo  {journal} {J. Phys. Soc. Japan}\ }\textbf {\bibinfo {volume}
  {45}},\ \bibinfo {pages} {786} (\bibinfo {year} {1978})}\BibitemShut
  {NoStop}%
\bibitem [{\citenamefont {Scalapino}(1969)}]{SuperconductivityVol1}%
  \BibitemOpen
  \bibfield  {author} {\bibinfo {author} {\bibfnamefont {D.}~\bibnamefont
  {Scalapino}},\ }\href@noop {} {\emph {\bibinfo {title} {{Superconductivity,
  Vol. 1}}}},\ edited by\ \bibinfo {editor} {\bibfnamefont {R.}~\bibnamefont
  {Parks}}\ (\bibinfo  {publisher} {Dekker, New York},\ \bibinfo {year}
  {1969})\ p.\ \bibinfo {pages} {449}\BibitemShut {NoStop}%
\bibitem [{\citenamefont {McMillan}(1968)}]{McMillan1968}%
  \BibitemOpen
  \bibfield  {author} {\bibinfo {author} {\bibfnamefont {W.~L.}\ \bibnamefont
  {McMillan}},\ }\href {\doibase 10.1103/PhysRev.167.331} {\bibfield  {journal}
  {\bibinfo  {journal} {Phys. Rev.}\ }\textbf {\bibinfo {volume} {167}},\
  \bibinfo {pages} {331} (\bibinfo {year} {1968})}\BibitemShut {NoStop}%
\bibitem [{\citenamefont {Allen}\ and\ \citenamefont
  {Dynes}(1975)}]{Allen1975}%
  \BibitemOpen
  \bibfield  {author} {\bibinfo {author} {\bibfnamefont {P.~B.}\ \bibnamefont
  {Allen}}\ and\ \bibinfo {author} {\bibfnamefont {R.~C.}\ \bibnamefont
  {Dynes}},\ }\href {\doibase 10.1103/PhysRevB.12.905} {\bibfield  {journal}
  {\bibinfo  {journal} {Phys. Rev. B}\ }\textbf {\bibinfo {volume} {12}},\
  \bibinfo {pages} {905} (\bibinfo {year} {1975})}\BibitemShut {NoStop}%
\bibitem [{\citenamefont {Morel}\ and\ \citenamefont
  {Anderson}(1962)}]{Morel1962}%
  \BibitemOpen
  \bibfield  {author} {\bibinfo {author} {\bibfnamefont {P.}~\bibnamefont
  {Morel}}\ and\ \bibinfo {author} {\bibfnamefont {P.~W.}\ \bibnamefont
  {Anderson}},\ }\href {\doibase 10.1103/PhysRev.125.1263} {\bibfield
  {journal} {\bibinfo  {journal} {Phys. Rev.}\ }\textbf {\bibinfo {volume}
  {125}},\ \bibinfo {pages} {1263} (\bibinfo {year} {1962})}\BibitemShut
  {NoStop}%
\bibitem [{\citenamefont {Methfessel}\ and\ \citenamefont
  {Paxton}(1989)}]{Methfessel1989}%
  \BibitemOpen
  \bibfield  {author} {\bibinfo {author} {\bibfnamefont {M.}~\bibnamefont
  {Methfessel}}\ and\ \bibinfo {author} {\bibfnamefont {A.~T.}\ \bibnamefont
  {Paxton}},\ }\href@noop {} {\bibfield  {journal} {\bibinfo  {journal} {Phys.
  Rev. B}\ }\textbf {\bibinfo {volume} {40}},\ \bibinfo {pages} {3616}
  (\bibinfo {year} {1989})}\BibitemShut {NoStop}%
\bibitem [{\citenamefont {Koretsune}\ and\ \citenamefont
  {Arita}(2016)}]{Koretsune2016}%
  \BibitemOpen
  \bibfield  {author} {\bibinfo {author} {\bibfnamefont {T.}~\bibnamefont
  {Koretsune}}\ and\ \bibinfo {author} {\bibfnamefont {R.}~\bibnamefont
  {Arita}},\ }\href {http://arxiv.org/abs/1610.09441} {\  (\bibinfo {year}
  {2016})},\ \Eprint {http://arxiv.org/abs/1610.09441} {arXiv:1610.09441}
  \BibitemShut {NoStop}%
\bibitem [{\citenamefont {Kawamura}\ \emph {et~al.}(2014)\citenamefont
  {Kawamura}, \citenamefont {Gohda},\ and\ \citenamefont
  {Tsuneyuki}}]{Kawamura2014}%
  \BibitemOpen
  \bibfield  {author} {\bibinfo {author} {\bibfnamefont {M.}~\bibnamefont
  {Kawamura}}, \bibinfo {author} {\bibfnamefont {Y.}~\bibnamefont {Gohda}}, \
  and\ \bibinfo {author} {\bibfnamefont {S.}~\bibnamefont {Tsuneyuki}},\ }\href
  {\doibase 10.1103/PhysRevB.89.094515} {\bibfield  {journal} {\bibinfo
  {journal} {Phys. Rev. B - Condens. Matter Mater. Phys.}\ }\textbf {\bibinfo
  {volume} {89}},\ \bibinfo {pages} {1} (\bibinfo {year} {2014})},\ \BibitemShut {NoStop}%
\bibitem [{\citenamefont {Carbotte}(1990)}]{Carbotte1990}%
  \BibitemOpen
  \bibfield  {author} {\bibinfo {author} {\bibfnamefont {J.~P.}\ \bibnamefont
  {Carbotte}},\ }\href {\doibase 10.1103/RevModPhys.62.1027} {\bibfield
  {journal} {\bibinfo  {journal} {Rev. Mod. Phys.}\ }\textbf {\bibinfo {volume}
  {62}},\ \bibinfo {pages} {1027} (\bibinfo {year} {1990})}\BibitemShut
  {NoStop}%
\bibitem [{\citenamefont {Akashi}\ \emph {et~al.}(2012)\citenamefont {Akashi},
  \citenamefont {Nakamura}, \citenamefont {Arita},\ and\ \citenamefont
  {Imada}}]{Akashi2012}%
  \BibitemOpen
  \bibfield  {author} {\bibinfo {author} {\bibfnamefont {R.}~\bibnamefont
  {Akashi}}, \bibinfo {author} {\bibfnamefont {K.}~\bibnamefont {Nakamura}},
  \bibinfo {author} {\bibfnamefont {R.}~\bibnamefont {Arita}}, \ and\ \bibinfo
  {author} {\bibfnamefont {M.}~\bibnamefont {Imada}},\ }\href {\doibase
  10.1103/PhysRevB.86.054513} {\bibfield  {journal} {\bibinfo  {journal} {Phys.
  Rev. B}\ }\textbf {\bibinfo {volume} {86}},\ \bibinfo {pages} {054513}
  (\bibinfo {year} {2012})}\BibitemShut {NoStop}%
\end{thebibliography}%

\end{document}